\documentclass[epj]{svjour}
\usepackage{graphicx}

\def\mb#1{\setbox0=\hbox{$#1$}\kern-.025em\copy0\kern-\wd0
\kern-0.05em\copy0\kern-\wd0\kern-.025em\raise.0233em\box0}

\begin{document}
   \title{Exact diffusion coefficient of self-gravitating Brownian particles in two dimensions}

 \author{P.H. Chavanis}

\institute{ Laboratoire de Physique Th\'eorique, Universit\'e Paul
Sabatier, 118 route de Narbonne 31062 Toulouse, France\\
\email{chavanis@irsamc.ups-tlse.fr}}

\titlerunning{}

   \date{To be included later }

   \abstract{We derive the exact expression of the diffusion
   coefficient of a self-gravitating Brownian gas in two
   dimensions. Our formula generalizes the usual Einstein relation for
   a free Brownian motion to the context of two-dimensional gravity. We
   show the existence of a critical temperature $T_{c}$ at which the
   diffusion coefficient vanishes. For $T<T_{c}$, the diffusion
   coefficient is negative and the gas undergoes gravitational
   collapse. This leads to the formation of a Dirac peak concentrating
   the whole mass in a finite time. We also stress that the critical
   temperature $T_{c}$ is different from the collapse temperature
   $T_{*}$ at which the partition function diverges. These quantities
   differ by a factor $1-1/N$ where $N$ is the number of particles in
   the system. We provide clear evidence of this difference by
   explicitly solving the case $N=2$. We also mention the analogy with
   the chemotactic aggregation of bacteria in biology, the formation
   of ``atoms'' in a two-dimensional (2D) plasma and the formation of
   dipoles or ``supervortices'' in 2D point vortex dynamics.  }

   \maketitle
%

\section{Introduction}
\label{sec_introduction}

Recently, there was a renewed interest for systems with long-range
interactions \cite{dauxois}. These systems have a very strange
thermodynamics and exhibit peculiar features that are very different
from those of more familiar systems with short-range interactions like
neutral gases and plasmas. One striking property of systems with
unshielded attractive long-range interactions is their ability to
self-organize spontaneously into large-scale coherent structures. Some
examples are provided by stars, globular clusters and galaxies in
astrophysics, jets and vortices (e.g. Jupiter's great red spot, gulf
stream,...) in two-dimensional geophysical flows, bacterial aggregates
in biology, clusters in the Hamiltonian Mean Field (HMF) model etc. As
a result of spatial inhomogeneity and non-extensivity, the ordinary
thermodynamic limit ($N\rightarrow +\infty$ with $N/V$ fixed) is
clearly irrelevant for these systems and must be reconsidered
\cite{pa1,pa2}. On the other hand, the statistical ensembles are
generically inequivalent and the choice of the relevant ensemble must
be addressed specifically. Therefore, systems with long-range
interactions are special and escape the ordinary rules of
thermodynamics. Among all types of systems with long-range
interactions, self-gravitating systems are probably the most
fundamental
\cite{paddy}. These systems have a complex thermodynamics and present
interesting phase transitions between ``gaseous'' and ``clustered''
states \cite{rev}. They can indeed undergo catastrophic collapse below a
critical energy $E_{c}$ in the microcanonical ensemble \cite{lbw} or
below a critical temperature $T_{c}$ in the canonical ensemble
\cite{aaiso} when gravitational attraction overcomes diffusive effects.

In astrophysics, the dynamics of self-gravitating systems is basically
described by the Newton equations where the acceleration of a particle
is equal to the gravitational force (per unit of mass) created by the
other particles \cite{bt,hut,saslaw}. These equations have a
Hamiltonian structure. For a large number of particles $N\gg 1$, we
cannot follow the motion of each particle in detail and we must have
recourse to statistical mechanics \cite{paddy,rev}.  An isolated
self-gravitating system is {\it conservative} (evolving at fixed
energy $E$) so that the proper statistical ensemble is the
microcanonical ensemble. This is the correct description of stellar
systems such as globular clusters and galaxies. The Hamiltonian
$N$-stars model has a very long history starting with Newton's {\it
Principia Mathematica} in 1687.

In a recent series of papers, Chavanis \& Sire [11-21]
have introduced and systematically studied a model of self-gravitating
Brownian particles. In this model, the particles interact
gravitationally but they also experience a friction force and a
stochastic force which mimick a coupling with a thermal bath of non
gravitational origin. Therefore, the basic equations of motion consist
in a set of $N$ coupled Langevin equations. This system is {\it
dissipative} (evolving at fixed temperature $T$) so that the proper
statistical ensemble to consider is the canonical ensemble. This
Brownian model could find applications in the process of planetesimal
formation in the solar nebula \cite{planetes}. In this context, the
dust particles experience a friction with the gas, a stochastic force
due to turbulence and, when the dust layer becomes dense enough,
self-gravity must be taken into account.  On the other hand, some
interesting analogies have been found between a self-gravitating
Brownian gas and the process of chemotaxis in biology
\cite{crrs,jeansbio}, the formation of large-scale vortices in 2D
turbulence \cite{houches,degrad} and the Bose-Einstein condensation in
the canonical ensemble \cite{bose}.  At a more academic level,
Brownian motion is a fundamental process in physics (pioneered by the
works of Einstein and Smoluchowski) and it is clearly of interest to
investigate the situation where $N$ Brownian particles are coupled by
a long-range potential of interaction, like the gravitational
interaction. These various arguments may justify the study of the
self-gravitating Brownian gas model.

Random walkers in interaction described by $N$ coupled Langevin
equations are also studied in soft matter physics to compute transport
properties of systems consisting in many interacting particles such as
supercooled liquids or the dynamics of colloids in solution
\cite{gotze,russel}.  In these examples, the potential of interaction is
short-range and the system is spatially homogeneous at
equilibrium. An interesting problem consists in determining the effective
diffusion coefficient of a particle of the system. This is a very
difficult problem and essentially perturbative approaches have been
developed. For example, Dean \& Lef\`evre
\cite{dl} consider a regime of weak coupling
(small $\beta$ with fixed $\beta\rho$) where diagrammatic expansions
can be carried out. Interestingly, their approach suggests that the
system will experience a glassy behaviour (identified by the
divergence of the relaxation time and the vanishing of the diffusion
coefficient) at some critical temperature $T_{c}$. However, since the
expansions are valid at high temperatures, it is not clear whether
their extrapolation to lower temperatures of order $T_{c}$ is
justified.

In this paper we show that, for the self-gravitating Brownian gas in two
dimensions, the diffusion
coefficient of a particle can be calculated {\it exactly} for any
temperature. It is given by the following expression
\begin{eqnarray}
\label{intro1}
D(T)=\frac{k_{B}T}{\xi m}\left (1-\frac{T_{c}}{T}\right ),
\end{eqnarray}
where $T_{c}$ is the critical temperature 
\begin{eqnarray}
\label{intro2}
k_{B}T_{c}=(N-1)\frac{Gm^{2}}{4}.
\end{eqnarray}
For $T\gg T_{c}$, when self-gravity becomes negligible, we
recover the Einstein relation \cite{einstein}. In that case, the
particles have a diffusive motion (corresponding to an evaporation of
the system) slightly modified by self-gravity. For $T=T_{c}$, the
diffusion coefficient vanishes and for $T<T_{c}$ it becomes negative
implying finite time collapse. In that case, the system forms a Dirac
peak containing the whole mass in a finite time. These different
regimes have been studied in detail in the mean field approximation by
analytically solving the Smoluchowski-Poisson system
\cite{sc,virial1}. Here, we present complementary results that are
valid beyond the mean field approximation. We stress that, contrary to
the case of systems with short-range interactions like those evocated
above \cite{gotze,russel,dl}, we calculate the diffusion coefficient
of a particle in a gravitational system that is out-of-equilibrium and
spatially inhomogeneous. The context is therefore very different from
the case of soft matter physics
\cite{gotze,russel,dl}. We also stress that the result
(\ref{intro1})-(\ref{intro2}) only holds in two-dimensional
gravity. Gravity is known to be critical in two dimensions because,
dimensionally, the gravitational potential $u=Gm^{2}\ln (r_{ij})$ does
not depend on the distance (the distance enters in a
dimensionless logarithmic factor). This is the intrinsic reason why
the Virial of the gravitational force and the diffusion coefficient
can be calculated exactly in 2D gravity.

The paper is organized as follows. In Sec. \ref{sec_mb}, we develop a
many-body theory of Brownian particles in interaction and derive exact
kinetic equations valid for an arbitrary binary potential of
interaction (Sec. \ref{sec_ek}). In Sec. \ref{sec_vt}, we derive the
general expression of the Virial theorem for a Brownian gas with
two-body interactions. When we consider the gravitational potential in
two dimensions (Sec. \ref{sec_vr}), the Virial theorem takes a very
simple form from which we can deduce the exact expression of the
diffusion coefficient (Sec. \ref{sec_dc}). The mean field
approximation is considered in Sec. \ref{sec_mf}.  In
Sec. \ref{sec_eos}, we obtain the exact equation of state of a
self-gravitating gas in two dimensions. This equation of state, first
derived in plasma physics for electric charges \cite{sp}, is
well-known. It is usually derived in the canonical ensemble from the
partition function \cite{salzberg,paddy}. Here, we show that the same
expression can be obtained in the microcanonical ensemble from the
density of states. We also generalize its expression by allowing the
particles to have different masses (or charges). In
Sec. \ref{sec_existence}, we investigate the existence of statistical
equilibrium for a two-dimensional self-gravitating system in the
canonical ensemble. We show that the partition function exists only
above a temperature $T_{*}$. This temperature differs from the
critical temperature $T_{c}$ appearing in the Virial theorem and in
the equation of state by a factor $1-1/N$ where $N$ is the number of
particles in the system. We provide clear evidence of this difference
by explicitly solving the case $N=2$ in Sec. \ref{sec_n2}. This
difference has been overlooked in the literature.
However, this is essentially a curiosity because, in the thermodynamic
limit $N\rightarrow +\infty$, the two temperatures coincide.

\section{Many-body theory}
\label{sec_mb}

\subsection{Exact kinetic equations}
\label{sec_ek}

In a space of dimension $d$, we consider a system of $N$ Brownian
particles confined within a domain (box) of volume $V$ and interacting via a
binary potential \cite{pa1,pa2}. Their dynamics is described by the
coupled Langevin equations
\begin{eqnarray}
\label{ek1}
\frac{d{\bf r}_{\alpha}}{dt}={\bf v}_{\alpha},
\end{eqnarray}
\begin{eqnarray}
\label{ek2}
\frac{d{\bf v}_{\alpha}}{dt}=-\xi_{\alpha} {\bf v}_{\alpha}-\frac{1}{m_{\alpha}}\nabla_{\alpha}U({\bf r}_{1},...,{\bf r}_{N})+\sqrt{2D'_{\alpha}}{\bf R}_{\alpha}(t),\nonumber\\
\end{eqnarray}
where $\xi_{\alpha}$ is a friction coefficient, $D'_{\alpha}$ a
diffusion coefficient and ${\bf R}_{\alpha}(t)$ a white noise
satisfying $\langle {\bf R}_{\alpha}(t)\rangle=0$ and $\langle
R_{\alpha,i}(t)R_{\beta,j}(t')\rangle
=\delta_{i,j}\delta_{\alpha,\beta}\delta(t-t')$ where $\alpha=1,...,N$
labels the particles and $i=1,...,d$ the coordinates of space. The particles interact via the potential
\begin{eqnarray}
\label{ek9}
U({\bf r}_{1},...,{\bf r}_{N})=\sum_{\alpha<\beta}m_{\alpha}m_{\beta} u(|{\bf r}_{\alpha}-{\bf r}_{\beta}|),
\end{eqnarray}
where $u(|{\bf r}-{\bf r}'|)$ is a binary potential of interaction
depending only on the absolute distance between the particles.  The
$N$-body Fokker-Planck equation describing the evolution of this
system is
\begin{eqnarray}
\label{ek3}
\frac{\partial P_{N}}{\partial t}+\sum_{\alpha=1}^{N}\left ({\bf v}_{\alpha}\cdot \frac{\partial P_{N}}{\partial {\bf r}_{\alpha}}+{\bf F}_{\alpha}\cdot \frac{\partial P_{N}}{\partial {\bf v}_{\alpha}} \right )\nonumber\\
=\sum_{\alpha=1}^{N}\frac{\partial}{\partial {\bf v}_{\alpha}}\cdot \left\lbrack D'_{\alpha}\frac{\partial P_{N}}{\partial {\bf v}_{\alpha}}+\xi_{\alpha}P_{N}{\bf v}_{\alpha}\right\rbrack,
\end{eqnarray}
where $P_{N}({\bf r}_{1},{\bf v}_{1},...,{\bf r}_{N},{\bf v}_{N},t)$
is the $N$-body distribution function and ${\bf
F}_{\alpha}=-\frac{1}{m_{\alpha}}\nabla_{\alpha}U$ is the force by
unit of mass experienced by particle $\alpha$.  In order to obtain the
canonical distribution at statistical equilibrium, we must impose the
Einstein relation
\begin{eqnarray}
\label{ek4}
D'_{\alpha}=\frac{\xi_{\alpha} k_{B}T}{m_{\alpha}}.
\end{eqnarray}
This shows that the temperature $T$ is a measure of the strength of
the stochastic force in Eq. (\ref{ek2}). Then, the stationary solution
of the Fokker-Planck equation (\ref{ek3}), cancelling independently \cite{pa2}
the advection term (l.h.s.) and the ``collision'' term (r.h.s.), is the
canonical distribution
\begin{eqnarray}
\label{ek5}
P_{N}({\bf r}_{1},{\bf v}_{1},...,{\bf r}_{N},{\bf v}_{N})=\frac{1}{Z(\beta)}e^{-\beta H({\bf r}_{1},{\bf v}_{1},...,{\bf r}_{N},{\bf v}_{N})},
\end{eqnarray}
where $H$ is the Hamiltonian
\begin{eqnarray}
\label{ek6}
H=\sum_{\alpha=1}^{N}m_{\alpha}\frac{v_{\alpha}^{2}}{2}+\sum_{\alpha<\beta}m_{\alpha}m_{\beta} u(|{\bf r}_{\alpha}-{\bf r}_{\beta}|).
\end{eqnarray}

In the overdamped limit where $\xi_{\alpha}\rightarrow +\infty$, we
can neglect the inertia of the particles in Eq. (\ref{ek2}) and we get
\begin{eqnarray}
\label{ek7}
\frac{d{\bf r}_{\alpha}}{dt}=-\mu_{\alpha}\nabla_{\alpha}U({\bf r}_{1},...,{\bf r}_{N})+\sqrt{2D_{\alpha}}{\bf R}_{\alpha}(t),
\end{eqnarray}
where we have introduced the mobility $\mu_{\alpha}=1/(\xi_{\alpha}
m_{\alpha})$ and the spatial diffusion coefficient
$D_{\alpha}=D'_{\alpha}/\xi_{\alpha}^{2}$. The $N$-body Fokker-Planck
equation describing the evolution of this system is
\begin{eqnarray}
\label{ek10}
\frac{\partial P_{N}}{\partial t}=\sum_{\alpha=1}^{N}\frac{\partial}{\partial {\bf r}_{\alpha}}\cdot \left\lbrack D_{\alpha}\frac{\partial P_{N}}{\partial {\bf r}_{\alpha}}+\mu_{\alpha}P_{N}\frac{\partial U}{\partial {\bf r}_{\alpha}}\right\rbrack,
\end{eqnarray}
where $P_{N}({\bf r}_{1},...,{\bf r}_{N},t)$ is the $N$-body
distribution in configuration space.  In order to obtain the canonical
distribution at statistical equilibrium, we must impose the Einstein
relation \footnote{For a multi-components system, a necessary  
condition for the Fokker-Planck equation (\ref{ek10}) to admit a stationary solution is that the ratio $\mu_{\alpha}/D_{\alpha}$ be independent on $\alpha$. This ratio is identified with the inverse temperature $\beta$.}
\begin{eqnarray}
\label{ek11}
\frac{\mu_{\alpha}}{D_{\alpha}}=\frac{1}{k_{B}T}\equiv \beta.
\end{eqnarray}
Using the expression of the mobility, the Einstein relation can be rewritten
\begin{eqnarray}
\label{ek12}
D_{\alpha}=\frac{k_{B}T}{\xi_{\alpha} m_{\alpha}}.
\end{eqnarray}
This expression can also be obtained from Eq. (\ref{ek4}).  Then, the
stationary solution of the Fokker-Planck equation (\ref{ek10}) is the
configurational part of the canonical distribution
\begin{eqnarray}
\label{ek13}
P_{N}({\bf r}_{1},...,{\bf r}_{N})=\frac{1}{Z(\beta)}e^{-\beta U({\bf r}_{1},...,{\bf r}_{N})}.
\end{eqnarray}
If we introduce the one and two-body probability distributions  
\begin{eqnarray}
\label{ek14}
P_{\alpha}({\bf r},t)=\int P_{N}({\bf r}_{1},...,{\bf r}_{N})\prod_{\gamma\neq \alpha}d{\bf r}_{\gamma}, 
\end{eqnarray}
\begin{eqnarray}
\label{ek15}
P_{\alpha,\beta}({\bf r},{\bf r}',t)=\int P_{N}({\bf r}_{1},...,{\bf r}_{N})\prod_{\gamma\neq \alpha,\beta}d{\bf r}_{\gamma},
\end{eqnarray}
we find from Eq. (\ref{ek10}) that the one-body distribution function
(\ref{ek14}) satisfies a kinetic equation of the form
\begin{eqnarray}
\label{ek16}
\frac{\partial P_{\alpha}}{\partial t}=\frac{\partial}{\partial {\bf r}}\cdot \biggl\lbrack D_{\alpha}\frac{\partial P_{\alpha}}{\partial {\bf r}}+\mu_{\alpha}\sum_{\beta\neq \alpha} m_{\alpha}m_{\beta}\nonumber\\
\times \int d{\bf r}' P_{\alpha,\beta}({\bf r},{\bf r}',t)\frac{\partial u}{\partial {\bf r}}(|{\bf r}-{\bf r}'|)\biggr\rbrack.
\end{eqnarray}
An alternative derivation of this equation is given in Appendix
\ref{sec_a}. This equation is exact and takes into account statistical
correlations encapsulated in the two-body distribution
(\ref{ek15}). As a result, this equation is not closed since it
involves a distribution of higher order. The complete hierarchy of
equations for the reduced distributions is given in
\cite{pa1,pa2}.

\subsection{The potential energy tensor}
\label{sec_pet}

We introduce the potential energy tensor \footnote{This is the usual $N$-body potential energy tensor \cite{bt} averaged over the noise.} 
\begin{eqnarray}
\label{pet1}
W_{ij}=-\sum_{\alpha}\sum_{\beta\neq \alpha} m_{\alpha}m_{\beta} \int d{\bf r}d{\bf r}' \nonumber\\
\times P_{\alpha,\beta}({\bf r},{\bf r}',t)x_{i}\frac{\partial u}{\partial x_{j}}(|{\bf r}-{\bf r}'|).
\end{eqnarray}
Since $u(|{\bf r}-{\bf r}'|)$ depends only on the absolute distance between the particles we get
\begin{eqnarray}
\label{pet2}
W_{ij}=-\sum_{\alpha\neq\beta} m_{\alpha}m_{\beta} \int d{\bf r}d{\bf r}' \nonumber\\
\times P_{\alpha,\beta}({\bf r},{\bf r}',t)x_{i}\frac{u'(|{\bf r}-{\bf r}'|)}{|{\bf r}-{\bf r}'|}(x_{j}-x_{j}').
\end{eqnarray}
Interchanging the dummy variables $\alpha,\beta$ and ${\bf r},{\bf r}'$, we have equivalently
\begin{eqnarray}
\label{pet3}
W_{ij}=\sum_{\alpha\neq\beta} m_{\alpha}m_{\beta} \int d{\bf r}d{\bf r}' \nonumber\\
\times P_{\alpha,\beta}({\bf r},{\bf r}',t)x'_{i}\frac{u'(|{\bf r}-{\bf r}'|)}{|{\bf r}-{\bf r}'|}(x_{j}-x_{j}').
\end{eqnarray}
Summing Eqs. (\ref{pet2}) and (\ref{pet3}), we obtain
\begin{eqnarray}
\label{pet4}
W_{ij}=-\frac{1}{2}\sum_{\alpha\neq\beta} m_{\alpha}m_{\beta} \int d{\bf r}d{\bf r}' \nonumber\\
\times P_{\alpha,\beta}({\bf r},{\bf r}',t)\frac{u'(|{\bf r}-{\bf r}'|)}{|{\bf r}-{\bf r}'|}(x_{i}-x_{i}')(x_{j}-x_{j}').
\end{eqnarray}
Under this form, the potential energy tensor is  manifestly symmetric
\begin{eqnarray}
\label{pet5}
W_{ij}=W_{ji}.
\end{eqnarray}
The trace of the potential energy tensor is
\begin{eqnarray}
\label{pet6}
W_{ii}=-\frac{1}{2}\sum_{\alpha\neq\beta} m_{\alpha}m_{\beta} \int d{\bf r}d{\bf r}' \nonumber\\
\times P_{\alpha,\beta}({\bf r},{\bf r}',t)u'(|{\bf r}-{\bf r}'|)|{\bf r}-{\bf r}'|.
\end{eqnarray}

\subsection{The exact Virial theorem}
\label{sec_vt}

We introduce the tensor of inertia \footnote{This expression
incorporates the friction coefficients $\xi_{\alpha}$. This generalization
proves to be necessary in order to obtain a closed expression of the Virial
theorem for a multi-components system. For identical particles,
the expression (\ref{vt1}) coincides with the usual tensor of inertia
of a $N$-body system \cite{bt} averaged over the noise.}
\begin{eqnarray}
\label{vt1}
I_{ij}(t)=\frac{1}{\xi}\sum_{\alpha}\int P_{\alpha}({\bf r},t)\xi_{\alpha}m_{\alpha}x_{i}x_{j}d{\bf r},
\end{eqnarray}
where $\xi\equiv \frac{1}{N}\sum_{\alpha}\xi_{\alpha}$ is the average
friction coefficient.  From Eq. (\ref{ek16}), we obtain
\begin{eqnarray}
\label{vt2}
\xi{\dot I}_{ij}=-\sum_{\alpha}\int d{\bf r} (x_{i}\delta_{kj}+x_{j}\delta_{ki})\biggl\lbrack k_{B}T\frac{\partial P_{\alpha}}{\partial x_{k}}\nonumber\\
+\sum_{\beta\neq \alpha} m_{\alpha}m_{\beta} \int d{\bf r}' P_{\alpha,\beta}({\bf r},{\bf r}',t)\frac{\partial u}{\partial x_{k}}(|{\bf r}-{\bf r}'|)\biggr\rbrack,
\end{eqnarray}
where we have used an integration by parts (the boundary term cancels
out since the current of diffusion vanishes on the box due to the
conservation of the normalization condition $\int
P_{\alpha}d{\bf r}=1$). On the other hand,
\begin{eqnarray}
\label{vt3}
-\int  (x_{i}\delta_{kj}+x_{j}\delta_{ki})\frac{\partial P_{\alpha}}{\partial x_{k}}d{\bf r}\nonumber\\
=2\delta_{ij}-\oint P_{\alpha}(x_{i}dS_{j}+x_{j}dS_{i}),
\end{eqnarray}
and
\begin{eqnarray}
\label{vt4}
-\sum_{\alpha}\int d{\bf r}\ (x_{i}\delta_{kj}+x_{j}\delta_{ki})\sum_{\beta\neq \alpha} m_{\alpha}m_{\beta} \nonumber\\
\times\int d{\bf r}' P_{\alpha,\beta}({\bf r},{\bf r}',t)\frac{\partial u}{\partial x_{k}}(|{\bf r}-{\bf r}'|)=2W_{ij}.
\end{eqnarray}
We thus obtain
\begin{eqnarray}
\label{vt5}
\xi {\dot I}_{ij}=2Nk_{B}T\delta_{ij}+2W_{ij}\nonumber\\
-\sum_{\alpha}k_{B}T\oint P_{\alpha}(x_{i}dS_{j}+x_{j}dS_{i}),
\end{eqnarray}
which is the general expression of the Virial theorem for a Brownian
gas of  particles in interaction in the overdamped limit.  Introducing
the local pressure 
\begin{eqnarray}
\label{vt6}
p({\bf r},t)=\sum_{\alpha}P_{\alpha}({\bf r},t)k_{B}T,
\end{eqnarray}
the Virial theorem can be rewritten
\begin{eqnarray}
\label{vt7}
\xi {\dot I}_{ij}=2Nk_{B}T\delta_{ij}+2W_{ij}\nonumber\\
-\oint p(x_{i}dS_{j}+x_{j}dS_{i}).
\end{eqnarray}
The scalar Virial theorem is obtained by contracting the indices leading to
\begin{eqnarray}
\label{vt8}
\frac{1}{2}\xi {\dot I}=dNk_{B}T+W_{ii}-\oint p {\bf r}\cdot d{\bf S},
\end{eqnarray}
where
\begin{eqnarray}
\label{vt9}
I(t)=\frac{1}{\xi}\sum_{\alpha}\int P_{\alpha}({\bf r},t)\xi_{\alpha}m_{\alpha}r^{2}d{\bf r},
\end{eqnarray}
is the generalized moment of inertia (including friction
coefficients). If the pressure is constant on the boundary of the
domain, we have
\begin{eqnarray}
\label{vt10q}
\oint p {\bf r}\cdot d{\bf S}=p_{b}\oint  {\bf r}\cdot d{\bf S}=p_{b}\int \nabla\cdot  {\bf r}\ d{\bf r}=d p_{b} V.
\end{eqnarray}
More generally, we introduce the notation
\begin{eqnarray}
\label{vt10}
P=\frac{1}{dV}\oint p {\bf r}\cdot d{\bf S},
\end{eqnarray}
which can be identified with a kinetic pressure. Then, the scalar
Virial theorem can be written
\begin{eqnarray}
\label{vt11}
\frac{1}{2}\xi {\dot I}=dNk_{B}T+W_{ii}-dPV.
\end{eqnarray}
At equilibrium ($\dot I=0$), we have
\begin{eqnarray}
\label{vt12}
dNk_{B}T+W_{ii}-dPV=0.
\end{eqnarray}
In the absence of interaction, we recover the perfect gas law
\begin{eqnarray}
\label{vt13}
PV=Nk_{B}T.
\end{eqnarray}

\subsection{The gravitational potential}
\label{sec_vr}

The gravitational potential in $d$ dimensions is given by
\begin{eqnarray}
\label{vr1}
u(\xi)=-\frac{1}{d-2}\frac{G}{\xi^{d-2}} \qquad (d\neq 2),
\end{eqnarray}
\begin{eqnarray}
\label{vr2}
u(\xi)=G\ln\xi \qquad (d=2),
\end{eqnarray}
where $G$ is the gravity constant (whose value depends on the
dimension of space).  The gravitational force is given by
\begin{eqnarray}
\label{vr3}
u'(\xi)=\frac{G}{\xi^{d-1}}.
\end{eqnarray}
Inserting this expression in Eq. (\ref{pet4}), we find that the
gravitational potential energy tensor reads
\begin{eqnarray}
\label{vr4}
W_{ij}=-\frac{G}{2}\sum_{\alpha\neq\beta} m_{\alpha}m_{\beta} \int d{\bf r}d{\bf r}' \nonumber\\
\times P_{\alpha,\beta}({\bf r},{\bf r}',t)\frac{(x_{i}-x_{i}')(x_{j}-x_{j}')}{|{\bf r}-{\bf r}'|^{d}},
\end{eqnarray}
and that the trace of the potential energy tensor is
\begin{eqnarray}
\label{vr5}
W_{ii}=-\frac{G}{2}\sum_{\alpha\neq\beta} m_{\alpha}m_{\beta} \int d{\bf r}d{\bf r}' \nonumber\\
\times P_{\alpha,\beta}({\bf r},{\bf r}',t)\frac{1}{|{\bf r}-{\bf r}'|^{d-2}}.
\end{eqnarray}
For $d\neq 2$, we have
\begin{eqnarray}
\label{vr6}
W_{ii}=(d-2)W,
\end{eqnarray}
where $W$ is the usual potential energy \cite{bt} averaged over the
noise. In that case, the scalar Virial theorem (\ref{vt11}) can be written
\begin{eqnarray}
\label{vr7}
\frac{1}{2}\xi {\dot I}=dNk_{B}T+(d-2)W-dPV.
\end{eqnarray}
For $d=2$ a ``miracle'' occurs and we get the simple exact result
\begin{eqnarray}
\label{vr7b}
W_{ii}=-\frac{G}{2}\sum_{\alpha\neq\beta}m_{\alpha}m_{\beta},
\end{eqnarray}
where we recall that
$\sum_{\alpha\neq\beta}=\sum_{\alpha=1}^{N}\sum_{\beta\neq
\alpha}$. We note that this expression, contrary to Eq. (\ref{vr6}),
does not depend on the configuration of the system but only on the
masses of the particles. We define a critical temperature
\begin{eqnarray}
\label{vr8}
k_{B}T_{c}=\frac{G\sum_{\alpha\neq\beta}m_{\alpha}m_{\beta}}{4N}.
\end{eqnarray}
For equal mass particles, we have
\begin{eqnarray}
\label{vr9}
k_{B}T_{c}=(N-1)\frac{Gm^{2}}{4}.
\end{eqnarray} 
Then, the scalar Virial theorem (\ref{vt11}) in $d=2$ can be written
\begin{eqnarray}
\label{vr10}
\frac{1}{4}\xi{\dot I}=Nk_{B}(T-T_{c})-PV.
\end{eqnarray}
At equilibrium, in a bounded domain, we get the exact equation of state
\begin{eqnarray}
\label{vr11}
PV=Nk_{B}(T-T_{c}),
\end{eqnarray}
where $P$ is the kinetic pressure (\ref{vt10}). In Sec. \ref{sec_eos},
we shall obtain this equation of state directly from the partition
function and deduce, by identification, that $P$ also represents the
thermodynamical pressure. Since $P\ge 0$, we conclude that a necessary
condition for the system to be at statistical equilibrium is that
$T\ge T_{c}$. For $T<T_{c}$, we cannot have statistical equilibrium
since the pressure would be negative. In fact, for $T<T_{c}$, we have
$\dot I\le \epsilon<0$ so that $I(t)=0$ at a finite time $t=t_{end}$
implying that the system forms a Dirac peak containing all the mass in
a finite time. Therefore, statistical equilibrium is not possible for
$T<T_{c}$. However, the condition $T\ge T_{c}$ does not guarantee
statistical equilbrium. We shall see in Sec. \ref{sec_existence} that
statistical equilibrium is possible only for $T\ge T_{*}$ where
$T_{*}$ is strictly larger than $T_{c}$. For $T_{c}<T<T_{*}$, the
system collapses (or has a peculiar temporal behaviour) so that 
Eq. (\ref{vr11}), corresponding to $\dot I=0$, is never satisfied.

In an infinite domain, assuming that initially the particles are in a
confined region of space, the pressure at infinity vanishes ($P=0$)
and the scalar Virial theorem (\ref{vr10}) reduces to
\begin{eqnarray}
\label{vr12}
\frac{1}{4}\xi{\dot I}=Nk_{B}(T-T_{c}).
\end{eqnarray}
This equation is readily integrated leading to
\begin{eqnarray}
\label{vr13}
{I}(t)=\frac{4Nk_{B}}{\xi}(T-T_{c})t+I(0).
\end{eqnarray}
For $T<T_{c}$, the moment of inertia  vanishes ($I=0$) at
\begin{eqnarray}
\label{vr14}
t_{end}=\frac{I(0)\xi}{4Nk_{B}(T_{c}-T)},
\end{eqnarray}
implying that the system forms a Dirac peak containing the whole mass
in a finite time $t_{end}$. For $T>T_{c}$, $I(t)\rightarrow +\infty$
for $t\rightarrow +\infty$ indicating that the system evaporates.

\subsection{The diffusion coefficient in $d=2$}
\label{sec_dc}

In this section, we restrict ourselves to a one component system. In
that case,
\begin{eqnarray}
\label{dc1}
I=\int\rho r^{2}d{\bf r},
\end{eqnarray}
is the usual moment of inertia with $\rho=NmP_{\alpha}$ for any
$\alpha=1,...,N$. The mean squared displacement of a particle (any) is
given by
\begin{eqnarray}
\label{dc2}
\langle r^{2}\rangle=\frac{\int\rho r^{2}d{\bf r}}{M}=\frac{I}{M}.
\end{eqnarray}
Thus, from Eq. (\ref{vr12}), we obtain
\begin{eqnarray}
\label{dc3}
\frac{d\langle r^{2}\rangle}{dt}=\frac{4k_{B}}{\xi m}(T-T_{c}),
\end{eqnarray}
where $T_{c}$ is given by Eq. (\ref{vr9}). After integration, we get
\begin{eqnarray}
\label{dc4}
\langle r^{2}\rangle=\frac{4k_{B}}{\xi m}(T-T_{c})t+\langle r^{2}\rangle_{0}.
\end{eqnarray}
The diffusion coefficient of the particle is defined by
\begin{eqnarray}
\label{dc5}
\langle r^{2}\rangle\sim 4D(T)t, \qquad (t\rightarrow +\infty).
\end{eqnarray}
Therefore, it is given by the exact expression
\begin{eqnarray}
\label{dc6}
D(T)=\frac{k_{B}}{\xi m}(T-T_{c}).
\end{eqnarray}
In the absence of gravity ($G=T_{c}=0$), or for high temperatures ($T\gg T_{c}$) where gravitational attraction becomes negligible with respect to thermal motion, we recover the usual expression of the diffusion coefficient given by the Einstein relation
\begin{eqnarray}
\label{dc7}
D=\frac{k_{B}T}{\xi m}.
\end{eqnarray}
However, for smaller temperatures, gravitational effects come into
play and the expression of the diffusion coefficient is
modified. Interestingly, there exists a critical temperature $T_{c}$
at which the diffusion coefficient vanishes \footnote{This is
physically different from the vanishing of the diffusion coefficient
in the case of colloids \cite{dl} which is due to the close packing of
the particles (steric hindrance) in the glassy phase (the particles
cannot move freely) while in the present situation we have a collapse due to
an attractive potential.}. For $T>T_{c}$ the diffusion coefficient is
positive so that the system evaporates.  For $T<T_{c}$ the diffusion
coefficient becomes negative implying finite time collapse to a Dirac
peak containing the whole mass in a time
\begin{eqnarray}
\label{dc8}
t_{end}=\frac{m\xi\langle r^{2}\rangle_{0}}{4k_{B}(T_{c}-T)}.
\end{eqnarray}
This time behaves like $(T_{c}-T)^{-1}$ for {\it any} $T<T_{c}$ and diverges
at the critical point $T_{c}$. At $T=T_{c}$ the Dirac peak is formed
in infinite time. These different regimes have been studied in
\cite{sc,banach,virial1} by solving the (mean field) Smoluchowski-Poisson
system in two dimensions.

\subsection{The mean field approximation}
\label{sec_mf}

The preceding results are exact for self-gravitating Brownian
particles in $d=2$ dimensions, whatever the number $N$ of
constituents. In particular, they take into account statistical
correlations. However, in most works on self-gravitating systems, one
usually considers a mean field approximation where the equations of
the problem can be simplified. In general, this approximation is valid
for $N\gg 1$. In this section, we briefly describe how the mean field
approximation can be implemented and how the results are modified.

The essence of the mean field approximation is to assume that the
two-body probability distribution  can be written as the product of two
one-body probability distributions according to
\begin{eqnarray}
\label{mf1}
P_{\alpha,\beta}({\bf r},{\bf r}',t)=P_{\alpha}({\bf r},t)P_{\beta}({\bf r}',t).
\end{eqnarray}
This approximation allows one to close the hierarchy of equations at the
level of Eq. (\ref{ek16}). For systems with long-range interactions,
it can be shown that this approximation is exact in a proper
thermodynamic limit $N\rightarrow +\infty$ with $\eta=\beta N u_{*}$
and $\epsilon=E/(u_{*}N^{2})$ fixed, where $u_{*}$ is the typical
value of the binary potential. For the gravitational potential in $d$
dimensions where $u_{*}=G m^2/R^{d-2}$, the thermodynamic limit
corresponds to $N\rightarrow +\infty$ in such a way that the
dimensionless temperature $\eta=\beta GMm/R^{d-2}$ and the
dimensionless energy $\epsilon=ER^{d-2}/GM^{2}$ are fixed. One can 
always rescale the quantities of the problem so that the
coupling constant ($G$ in gravity) scales like $u_{*}\sim 1/N$
while $E\sim N$, $T\sim 1$ and $V\sim 1$
\cite{pa1}. In this limit, the factorization (\ref{mf1}) is valid 
up to terms of order $1/N$.  Since we consider a large number limit,
we can also extend the sum in Eq. (\ref{ek16}) over all the
particles. This yields
\begin{eqnarray}
\label{mf2}
\frac{\partial P_{\alpha}}{\partial t}=\frac{\partial}{\partial {\bf r}}\cdot \biggl\lbrack D_{\alpha}\frac{\partial P_{\alpha}}{\partial {\bf r}}+\mu_{\alpha} m_{\alpha} P_{\alpha}({\bf r},t) \nonumber\\
\times  \nabla \sum_{\beta}m_{\beta} \int d{\bf r}' P_{\beta}({\bf r}',t)u(|{\bf r}-{\bf r}'|)\biggr\rbrack.
\end{eqnarray}
If we introduce the density
\begin{eqnarray}
\label{mf3}
\rho({\bf r},t)=\sum_{\alpha} P_{\alpha}({\bf r},t) m_{\alpha},
\end{eqnarray} 
we get
\begin{eqnarray}
\label{mf4}
\frac{\partial P_{\alpha}}{\partial t}=\frac{\partial}{\partial {\bf r}}\cdot \biggl\lbrack D_{\alpha}\frac{\partial P_{\alpha}}{\partial {\bf r}}+\mu_{\alpha} m_{\alpha} P_{\alpha}({\bf r},t) \nonumber\\
\times  \nabla \int d{\bf r}' \rho({\bf r}',t)u(|{\bf r}-{\bf r}'|)\biggr\rbrack.
\end{eqnarray}
This equation can be rewritten in the form of a mean field
Smoluchowski equation
\begin{eqnarray}
\label{mf5}
\frac{\partial P_{\alpha}}{\partial t}=\frac{\partial}{\partial {\bf r}}\cdot \biggl\lbrack D_{\alpha}\frac{\partial P_{\alpha}}{\partial {\bf r}}+\mu_{\alpha} m_{\alpha} P_{\alpha}({\bf r},t) \nabla \Phi\biggr\rbrack,
\end{eqnarray}
where the potential is determined by the density according to 
\begin{eqnarray}
\label{mf6}
\Phi({\bf r},t)=\int \rho({\bf r}',t)u(|{\bf r}-{\bf r}'|) d{\bf r}'.
\end{eqnarray}
For the gravitational potential, the preceding equation is equivalent to the Poisson equation
\begin{eqnarray}
\label{mf7}
\Delta\Phi=S_{d}G\rho.
\end{eqnarray}
Therefore, in the mean field approximation, we have to solve the
multi-components Smoluchowski-Poisson system (\ref{mf5})-(\ref{mf7}). The steady state of Eq. (\ref{mf5}) is the mean field Boltzmann distribution
\begin{eqnarray}
\label{mf8}
P_{\alpha}({\bf r})=A_{\alpha}e^{-\beta m_{\alpha}\Phi({\bf r})}.
\end{eqnarray}
The single component Smoluchowski-Poisson system has been studied in
[11-17,19-21] and the two components Smoluchowski-Poisson system has
been studied in \cite{sopik}.

To establish the expression of the Virial theorem in the mean field
approximation, we can use a procedure similar to the one developed in
Sec. \ref{sec_vt}. The only change is the factorization (\ref{mf1}) and the
replacement of $\sum_{\beta\neq \alpha}$ by $\sum_{\beta}$. Thus, the
previous relations remain valid provided that $W_{ij}$ is replaced by
\begin{eqnarray}
\label{mf9}
W_{ij}^{MF}=-\int \rho({\bf r},t)x_{i}\frac{\partial \Phi}{\partial x_{j}},
\end{eqnarray} 
where $\Phi({\bf r},t)$ is given by Eq. (\ref{mf6}). In particular, for the gravitational interaction we have
\begin{eqnarray}
\label{mf10}
W_{ij}^{MF}=-\frac{G}{2}\int d{\bf r}d{\bf r}' \nonumber\\
\times \rho({\bf r},t)\rho({\bf r}',t)\frac{(x_{i}-x_{i}')(x_{j}-x_{j}')}{|{\bf r}-{\bf r}'|^{d}},
\end{eqnarray}
and the trace of the potential energy tensor is
\begin{eqnarray}
\label{mf11}
W_{ii}^{MF}=-\frac{G}{2} \int d{\bf r}d{\bf r}'\frac{\rho({\bf r},t)\rho({\bf r}',t)}{|{\bf r}-{\bf r}'|^{d-2}}.
\end{eqnarray}
For $d\neq 2$, 
\begin{eqnarray}
\label{mf12}
W_{ii}^{MF}=(d-2)W^{MF},
\end{eqnarray}
where $W^{MF}$ is the mean field potential energy 
\begin{eqnarray}
\label{mf13}
W^{MF}=\frac{1}{2} \int \rho\Phi d{\bf r},
\end{eqnarray}
with 
\begin{eqnarray}
\label{mf14}
\Phi({\bf r},t)=-\frac{G}{d-2} \int \frac{\rho({\bf r},t)}{|{\bf r}-{\bf r}'|^{d-2}}d{\bf r}'.
\end{eqnarray}
For $d=2$, we get 
\begin{eqnarray}
\label{mf15}
W_{ii}^{MF}=-\frac{G}{2} \int d{\bf r}d{\bf r}'\rho({\bf r},t)\rho({\bf r}',t)=-\frac{GM^{2}}{2}.
\end{eqnarray}
Therefore, in the mean field approximation, the  scalar Virial theorem of a self-gravitating Brownian gas in $d=2$ can be written as in Eq. (\ref{vr10}) with the critical temperature
\begin{eqnarray}
\label{mf16}
k_{B}T_{c}^{MF}=\frac{GM^{2}}{4N}.
\end{eqnarray}
For equal mass particles it reduces to
\begin{eqnarray}
\label{mf17}
k_{B}T_{c}^{MF}=\frac{GNm^{2}}{4}.
\end{eqnarray}
These expressions can be directly obtained from Eqs. (\ref{vr8}) and
(\ref{vr9}) by replacing $\sum_{\alpha\neq\beta}$ by
$\sum_{\alpha}\sum_{\beta}$ in Eq. (\ref{vr8}) or by replacing $N-1$
by $N$ in Eq. (\ref{vr9}) since the mean field approximation is valid
for $N\gg1 $. We note that the mean field results are relatively close
to the exact results even for a moderate number of particles. Equation
(\ref{vr10}) is always valid and finite $N$ effects just slightly
shift the critical temperature $T_{c}$. Since $N-1=N(1-1/N)$, the
correction is of order $1/N$, which is precisely the domain of
validity of the factorization hypothesis (\ref{mf1}) as shown in
\cite{pa1,pa2} at a more general level. This corroborates the
observation that the mean field approximation provides a good
description of systems with long-range interactions such as
self-gravitating systems.

\section{The exact equation of state}
\label{sec_eos}

In this section, we derive the exact equation of state of a
self-gravitating gas in two dimensions. We extend the derivation given
by Salzberg \cite{salzberg} and Padmanabhan \cite{paddy} in two
respects: 1. we consider a multi-components system while the previous
authors assume that the particles have the same mass 2. we treat both
the canonical and the microcanonical ensembles while the previous
authors only consider the canonical ensemble. For comparison, we also
discuss the case of a two-dimensional plasma made of electric charges
\cite{sp}.

\subsection{Canonical approach}
\label{sec_ca}

For the gravitational interaction in two dimensions
\begin{eqnarray}
\label{ca1}
U({\bf r}_{1},...,{\bf r}_{N})=G\sum_{i<j}m_{i}m_{j}\ln |{\bf r}_{i}-{\bf r}_{j}|,
\end{eqnarray}
the configurational part of the partition function in the canonical
ensemble is given by
\begin{eqnarray}
\label{ca2}
Z(\beta,V)=\int e^{-\beta G\sum_{i<j}m_{i}m_{j}\ln |{\bf r}_{i}-{\bf r}_{j}|}\prod_{k=1}^{N}d{\bf r}_{k}.
\end{eqnarray}
To avoid the divergence of the partition function at large distances,
we assume that the system is enclosed within a box of radius $R$. The
following calculations also assume implicitly that the partition
function converges at small distances. The existence of statistical
equilibrium states will be discussed in
Sec. \ref{sec_existence}. Using a trick due to Salzberg
\cite{salzberg}, we set ${\bf y}={\bf r}/R$. Then, the partition
function can be rewritten
\begin{eqnarray}
\label{ca3}
Z(\beta,V)=R^{2N}e^{-\beta G\ln R\sum_{i<j}m_{i}m_{j}}\nonumber\\
\int e^{-\beta G\sum_{i<j}m_{i}m_{j}\ln |{\bf y}_{i}-{\bf y}_{j}|}\prod_{k=1}^{N}d{\bf y}_{k},
\end{eqnarray} 
where the last integral is now {\it independent} on $R$. From this
expression, we find that
\begin{eqnarray}
\label{ca4}
\frac{\partial Z}{\partial R}=\frac{2N}{R}\left\lbrack 1-\frac{\beta G}{2N}\sum_{i<j}m_{i}m_{j}\right\rbrack Z(\beta,R).
\end{eqnarray} 
The thermodynamic pressure is defined by 
\begin{eqnarray}
\label{ca5}
P=\frac{1}{\beta}\frac{\partial \ln Z}{\partial V}, \qquad V=\pi R^{2}.
\end{eqnarray} 
From Eqs. (\ref{ca5}) and (\ref{ca4}), we obtain the exact equation of state of a two-dimensional multi-components self-gravitating gas
\begin{eqnarray}
\label{ca6}
P=\frac{N}{\beta V}\left\lbrack 1-\frac{\beta G}{2N}\sum_{i<j}m_{i}m_{j}\right\rbrack.
\end{eqnarray} 
This can be written
\begin{eqnarray}
\label{ca7}
PV=Nk_{B}(T-T_{c}),
\end{eqnarray}
with the critical temperature
\begin{eqnarray}
\label{ca8}
k_{B}T_{c}=\frac{G\sum_{i\neq j}m_{i}m_{j}}{4N}.
\end{eqnarray}
For equal mass particles, we have
\begin{eqnarray}
\label{ca9}
k_{B}T_{c}=(N-1)\frac{Gm^{2}}{4}.
\end{eqnarray}
This returns the equation of state (\ref{vr11}) obtained by the
kinetic approach. Note that for a single species system in the
mean field approximation, the equation of state (\ref{ca7}) can also
be obtained by solving the Boltzmann-Poisson equation and computing
the pressure at the edge of the box \cite{klb}.

\subsection{Microcanonical approach}
\label{sec_ma}

The Hamiltonian of a self-gravitating system in two dimensions is 
\begin{eqnarray}
\label{ma1}
H=\sum_{i=1}^{N}\frac{1}{2}m_{i}v_{i}^{2}+G\sum_{i<j}m_{i}m_{j}\ln |{\bf r}_{i}-{\bf r}_{j}|.
\end{eqnarray}
The density of states in the microcanonical ensemble is defined by
\begin{eqnarray}
\label{ma2}
g(E,R)=\int \delta(E-H)\prod_{k=1}^{N}d{\bf r}_{k}d{\bf v}_{k}.
\end{eqnarray}
With the transformation ${\bf y}={\bf r}/R$, it can be written
\begin{eqnarray}
\label{ma3}
g(E,R)=R^{2N}g(E',1),
\end{eqnarray}  
with
\begin{eqnarray}
\label{ma4}
E'=E-G\ln R\sum_{i<j}m_{i}m_{j}.
\end{eqnarray} 
The entropy is defined by
\begin{eqnarray}
\label{ma5}
S(E,R)=k_{B}\ln g(E,R).
\end{eqnarray}  
According to Eq. (\ref{ma3}), we have
\begin{eqnarray}
\label{ma6}
S(E,R)=2Nk_{B}\ln R+S(E',1).
\end{eqnarray} 
The temperature and the pressure are given by
\begin{eqnarray}
\label{ma7}
\frac{1}{T}=\left (\frac{\partial S}{\partial E}\right )_{N,V}, \qquad P=T\left (\frac{\partial S}{\partial V}\right )_{N,E},
\end{eqnarray} 
with $V=\pi R^{2}$. From Eqs. (\ref{ma6}), (\ref{ma4}) and (\ref{ma7})-a we find that
\begin{eqnarray}
\label{ma8}
\left (\frac{\partial S}{\partial R}\right )_{N,E}= \frac{2Nk_{B}}{R}-\frac{G}{TR}\sum_{i<j}m_{i}m_{j}.
\end{eqnarray} 
Using Eq. (\ref{ma7})-b, we obtain 
\begin{eqnarray}
\label{ma9}
PV=Nk_{B}\left \lbrack T(E)-\frac{G}{2Nk_{B}}\sum_{i<j}m_{i}m_{j}\right \rbrack.
\end{eqnarray} 
This returns the expression (\ref{ca7}) obtained in the canonical
ensemble with the critical temperature (\ref{ca8}) and (\ref{ca9}). Therefore, the equations of state coincide in the two ensembles for
any number of particles. This result was not obvious at first
sights. For two-dimensional self-gravitating systems, the
microcanonical and canonical ensembles are equivalent at the
thermodynamic limit with $N\rightarrow +\infty$ \footnote{This is
because the series of equilibria of a two-dimensional self-gravitating
gas does not present turning points contrary to 3D self-gravitating
systems; see, e.g., \cite{sc}.} but they are not equivalent for finite
values of $N$. For example, we shall explicitly show in
Sec. \ref{sec_n2} that the caloric curves for $N=2$ calculated in the
microcanonical and canonical ensembles differ. Yet, the equation of
state is the same for any $N$.

\subsection{The case of electric charges}
\label{sec_ec}

The case of electric charges is obtained by taking $G=-1$ and by
making the substitution $m\leftrightarrow q$. Therefore, the energy of
interaction in two dimensions reads
\begin{eqnarray}
\label{ec1}
U({\bf r}_{1},...,{\bf r}_{N})=-\sum_{i<j}q_{i}q_{j}\ln |{\bf r}_{i}-{\bf r}_{j}|.
\end{eqnarray}
The previous results remain valid for electric charges instead of point masses, provided that we make the above mentioned substitutions. In particular, Eqs. (\ref{ca6}) and (\ref{ma9}) are now replaced by
\begin{eqnarray}
\label{ec2}
P=\frac{Nk_{B}T}{V}\left\lbrack 1+\frac{\beta }{2N}\sum_{i<j}q_{i}q_{j}\right\rbrack.
\end{eqnarray} 
Using the condition of electroneutrality
\begin{eqnarray}
\label{ec3}
\sum_{i=1}^{N}q_{i}=0,
\end{eqnarray} 
we find that
\begin{eqnarray}
\label{ec4}
\sum_{i<j}q_{i}q_{j}=\frac{1}{2}\sum_{i\neq j}q_{i}q_{j}\nonumber\\
=\frac{1}{2}\sum_{i=1}^{N}q_{i}\left (\sum_{j=1}^{N}q_{j}-q_{i}\right )=-\frac{1}{2}\sum_{i=1}^{N}q_{i}^{2}.
\end{eqnarray} 
Therefore, the equation of state of a neutral two-dimensional plasma
can be written
\begin{eqnarray}
\label{ec5}
PV=Nk_{B}(T-T_{c}),
\end{eqnarray}
with the critical temperature
\begin{eqnarray}
\label{ec6}
k_{B}T_{c}=\frac{\sum_{i=1}^{N} q_{i}^{2}}{4N}.
\end{eqnarray}
If the plasma consists in $N/2$ charges $+e$ and $N/2$ charges $-e$, we get 
\begin{eqnarray}
\label{ec7}
k_{B}T_{c}=\frac{e^2}{4}.
\end{eqnarray}
These results have been first derived by Salzberg \& Prager \cite{sp}
in the canonical ensemble. The calculations of Sec. \ref{sec_ma} show
that they can also be derived in the microcanonical ensemble. Since
$P\ge 0$, a necessary condition for the system to be at statistical
equilibrium is that $T\ge T_{c}$. However, this condition does not
guarantee statistical equilibrium. We shall see in the sequel that
statistical equilibrium exists only for $T\ge T_{*}=2T_{c}$. For
$T<T_{*}$ there is no equilibrium state and the system collapses. This
leads to the formation of $N/2$ pairs $(+,-)$ corresponding to non
interacting ``atoms''. By contrast, for $T>T_{*}$ the system is fully
ionized. For comparison, a gas of self-gravitating Brownian particles
collapses to a single Dirac peak containing the $N$ particles for
$T<T_{*}={N\over N-1}T_{c}$ while it remains diffuse for
$T>T_{*}$. Since the collapse in plasma physics leads to the formation
of individual pairs, we understand qualitatively why the collapse
temperature $k_{B}T_{*}={e^{2}}/{2}$ does not depend on $N$ (it
corresponds to the condition to form {\it one} pair). By contrast,
since the collapse in gravity leads to the formation of a single Dirac
peak containing all the mass, the collapse temperature
$T_{*}={NGm^{2}}/{4}$ depends on $N$ (it corresponds to the condition
to form a cluster of $N$ particles). Finally, it is easy to determine
the critical temperature of a non neutral 2D plasma consisting in
$N_{+}$ charges $+e$ and $N_{-}$ charges $-e$. From, Eqs. (\ref{ec2})
and (\ref{ec5}), we find
\begin{eqnarray}
\label{ec7net}
k_{B}T_{c}=\frac{e^2}{4}\left\lbrack 1-\frac{(N_{+}-N_{-})^{2}}{N}\right\rbrack.
\end{eqnarray}

\section{Existence of statistical equilibrium in the canonical ensemble}
\label{sec_existence}

For a finite two-dimensional self-gravitating system, there exists
statistical equilibrium states for any value of the energy in the
microcanonical ensemble (Hamiltonian systems). Indeed, if the system
is enclosed within a box, the density of states $g(E)$ is convergent
for any $E$. By contrast, in the canonical ensemble (Brownian systems)
equilibrium states exist only for sufficiently high temperatures
$T>T_{*}$. Although the existence of a collapse temperature $T_{*}$ is
well-known, there is some ambiguity in the literature concerning its
precise value. In particular, we will show that the collapse
temperature $T_{*}$ does not exactly coincide with the critical
temperature $T_{c}$ introduced previously. We will also show that the
collapse temperature $T_{*}$ is difficult to determine for a
multi-components system while it takes a simple expression when the
particles have the same mass $m$.

\subsection{Statistical equilibrium state for $T>T_{1}$}
\label{sec_ex}

In this section, using the arithmetic-geometric mean inequality, we
show that the partition function $Z(\beta)$ is convergent for
$T>T_{1}$. We extend the method developed by Kiessling
\cite{kiessling} to the case of a multi-components system. First, we
note that
\begin{eqnarray}
\label{ex1}
e^{-\beta G\sum_{i<j}m_{i}m_{j}\ln |{\bf r}_{i}-{\bf r}_{j}|}\nonumber\\
=e^{-\frac{1}{2}\beta G\sum_{i}\sum_{j\neq i}m_{i}m_{j}\ln |{\bf r}_{i}-{\bf r}_{j}|}\nonumber\\
=\prod_{i} e^{-\frac{1}{2}\beta G\sum_{j\neq i}m_{i}m_{j}\ln |{\bf r}_{i}-{\bf r}_{j}|}\nonumber\\
=\prod_{i}\left\lbrack e^{-\frac{N}{2}\beta G\sum_{j\neq i}m_{i}m_{j}\ln |{\bf r}_{i}-{\bf r}_{j}|}\right\rbrack^{1/N}.
\end{eqnarray}
Now, by the arithmetic-geometric mean inequality
\begin{eqnarray}
\label{ex2}
\frac{1}{N}\sum_{i}a_{i}\ge \prod_{i}a_{i}^{1/N},
\end{eqnarray}
we get
\begin{eqnarray}
\label{ex3}
e^{-\beta G\sum_{i<j}m_{i}m_{j}\ln |{\bf r}_{i}-{\bf r}_{j}|}\nonumber\\
\le \frac{1}{N}\sum_{i} e^{-\frac{N}{2}\beta G\sum_{j\neq i}m_{i}m_{j}\ln |{\bf r}_{i}-{\bf r}_{j}|}\nonumber\\
= \frac{1}{N}\sum_{i}\prod_{j\neq i} e^{-\frac{N}{2}\beta G m_{i}m_{j}\ln |{\bf r}_{i}-{\bf r}_{j}|}.
\end{eqnarray}
From this inequality, we find that the configurational part of the partition function satisfies
\begin{eqnarray}
\label{ex4}
Z\le \frac{1}{N}\sum_{i} \int d{\bf r}_{i}\prod_{j\neq i}\int d{\bf r}_{j} e^{-\frac{N}{2}\beta G m_{i}m_{j}\ln |{\bf r}_{i}-{\bf r}_{j}|}.\nonumber\\
\end{eqnarray}
The integrand is maximum for ${\bf r}_{i}={\bf 0}$ yielding
\begin{eqnarray}
\label{ex5}
Z\le \frac{1}{N}\sum_{i} \int d{\bf r}_{i}\prod_{j\neq i}\int  e^{-\frac{N}{2}\beta G m_{i}m_{j}\ln x}\ d{\bf x}\nonumber\\
=\frac{\pi R^{2}}{N}\sum_{i}\prod_{j\neq i}\int  x^{-\frac{N}{2}\beta G m_{i}m_{j}}\ d{\bf x}\nonumber\\
=\frac{2\pi^{2} R^{2}}{N}\sum_{i}\prod_{j\neq i}\int_{0}^{R}  x^{1-\frac{N}{2}\beta G m_{i}m_{j}}dx.
\end{eqnarray}
If 
\begin{eqnarray}
\label{ex6}
k_{B}T>\frac{NGm_{i}m_{j}}{4}
\end{eqnarray}
for all $i,j$ with $i\neq j$, then all the integrals converge and the partition function is finite. Therefore, a sufficient condition for the existence of statistical equilibrium is that $T>T_{1}$ with 
\begin{eqnarray}
\label{ex7}
k_{B}T_{1}=\frac{NGm_{I}m_{II}}{4},
\end{eqnarray}
where $I$ and $II$ denote the two most massive particles in the
system. In that case, we obtain the following bound on the partition
function
\begin{eqnarray}
\label{ex8}
Z\le \frac{\pi R^{2}}{N}\sum_{i}\prod_{j\neq i}\frac{\pi R^{2-\frac{N}{2}\beta G m_{i}m_{j}}}{1-\frac{N}{4}\beta G m_{i}m_{j}}.
\end{eqnarray}
For equal mass particles, a sufficient condition for the existence of statistical equilibrium is that $T>T_{1}$ with 
\begin{eqnarray}
\label{ex9}
k_{B}T_{1}=\frac{NGm^{2}}{4}.
\end{eqnarray}
 In that case, we recover the bound on the partition function given by Kiessling \cite{kiessling}:
\begin{eqnarray}
\label{ex10}
Z\le \pi R^{2}\left (\frac{\pi R^{2-\frac{N}{2}\beta G m^{2}}}{1-\frac{N}{4}\beta G m^{2}}\right )^{N-1}.
\end{eqnarray}

\subsection{Collapse for $T<T_{2}$}
\label{sec_co}

We now show that the partition function is divergent for $T<T_{2}$. We first rewrite the configurational partition function in the form
\begin{eqnarray}
\label{co1}
Z=\int d{\bf r}_{1}...d{\bf r}_{N} \prod_{i\neq j} \frac{1}{|{\bf r}_{i}-{\bf r}_{j}|^{\frac{1}{2}\beta G m_{i}m_{j}}}.
\end{eqnarray}
We select a particle (allowed to move over the whole box) and approach the $N-1$ other particles at a distance $\epsilon$ of the first \cite{kiessling,caglioti}. The contribution of this configuration to the partition function behaves like
\begin{eqnarray}
\label{co2}
Z\sim R^{2}\epsilon^{2(N-1)} \prod_{i\neq j} \left (\frac{1}{\epsilon}\right )^{\frac{1}{2}\beta G m_{i}m_{j}}\nonumber\\
= R^{2} \left (\frac{1}{\epsilon}\right )^{\sum_{i\neq j}\frac{1}{2}\beta G m_{i}m_{j}-2(N-1)}.
\end{eqnarray}
Considering now the limit $\epsilon\rightarrow 0$, we see that the partition function diverges if  $T<T_{2}$ with
\begin{eqnarray}
\label{co3}
k_{B}T_{2}=\frac{G\sum_{i\neq j} m_{i}m_{j}}{4(N-1)}.
\end{eqnarray}
For equal mass particles, we find that 
\begin{eqnarray}
\label{co4}
k_{B}T_{2}=N\frac{Gm^{2}}{4}.
\end{eqnarray}

\subsection{The collapse temperature $T_{*}$}
\label{sec_tt}

In conclusion, the partition function converges for $T>T_{1}$ and 
diverges for $T<T_2$. For $T_{2}<T<T_{1}$, we have no result in the
general case. However,  for equal mass
particles,  we have $T_{1}=T_{2}=T_{*}$ with
\begin{eqnarray}
\label{tt1}
k_{B}T_{*}=N\frac{Gm^{2}}{4}.
\end{eqnarray}
Therefore, the partition function converges for $T>T_{*}$ and diverges
for $T<T_*$. For $T>T_{*}$, we have a diffuse gas and for
$T<T_{*}$ the most probable distribution is a Dirac peak containing all the particles. We note that $T_{*}$ {\it differs} by a factor $1-1/N$
from the critical temperature $T_{c}$ appearing in the equation of
state (\ref{ca7}). Indeed,
\begin{eqnarray}
\label{tt1bg}
T_{*}={N\over N-1}T_{c}.
\end{eqnarray}
This point has been overlooked in the
literature. In Sec. \ref{sec_n2}, we shall provide a clear evidence of
this difference by considering explicity the case $N=2$ where the
partition function can be calculated exactly. Finally, we note that the pressure (\ref{ca7}) at the collapse
temperature $T_{*}$ is equal to
\begin{eqnarray}
\label{tt1bgg}
PV=k_{B}T_{*}, \qquad (T=T_{*}).
\end{eqnarray}
This corresponds to the ideal pressure due to  $N_{*}=1$ effective particle
(the Dirac peak) containing the whole mass when the system has collapsed.

For a multi-components system, the collapse temperature $T_{*}$
seems difficult to obtain. The case of a two-components system has
been considered  by 
\cite{sopik} in the mean field approximation 
and an explicit expression of the collapse temperature has been
obtained in that case.

\subsection{The case of 2D plasmas}
\label{sec_ttpla}

For self-gravitating systems, the partition function diverges for
$T\le T_{*}$ due to the collapse of {\it all} the particles at the
same point. Thus, below $T_{*}$, the most probable structure is a
Dirac peak containing the whole mass. By using an argument similar to
that of Eq. (\ref{co2}), we see that if we concentrate only a fraction of
the particles (e.g. a pair), we obtain a smaller collapse temperature,
so that this configuration is less favorable.

By contrast, for neutral plasmas with $N/2$ charges $+e$ and $N/2$
charges $-e$, the divergence of the partition function corresponds to
the formation (collapse) of individual pairs $(+,-)$. Therefore, we
can obtain the collapse temperature by considering the
statistical mechanics of only {\it one} pair ($N=2$). This is done it the
next section and the analysis yields
\begin{eqnarray}
\label{ddc15}
k_{B}T_{*}={e^{2}\over 2}.
\end{eqnarray}   
Therefore, the partition function of the $N$-body system converges for
$T>T_{*}$ and diverges for $T<T_*$. For $T>T_{*}$, we have a fully
ionized gas and for $T<T_{*}$ the most probable distribution is $N/2$
pairs $(+,-)$. We note that $T_{*}$ {\it differs} by a factor $2$ from
the critical temperature $T_{c}$ appearing in the equation of state
(\ref{ec5}). Indeed,
\begin{eqnarray}
\label{ddc15b}
T_{*}=2T_{c}.
\end{eqnarray}  
Since the equilibrium states exists only for $T>T_{*}$, the
equation of state (\ref{ec5}) is valid only above $T_{*}$. For
$T<T_{*}$ (including the region $T_{c}<T<T_{*}$), the partition
function diverges due to the collapse of pairs of point particles of
opposite sign. Finally, we note that the pressure (\ref{ec5}) at the collapse
temperature $T_{*}$ is equal to
\begin{eqnarray}
\label{ddc15c}
PV=\frac{1}{2}N k_{B}T_{*}, \qquad (T=T_{*}).
\end{eqnarray}
This is the pressure created by an {\it ideal} gas of $N_{*}=N/2$
non-interacting pairs $(+,-)$. Finally, for $T<T_{c}$, the Virial
theorem shows that the $N_{*}$ pairs collapse to a {\it single} point
in a finite time (if the motion of the charges in the plasma is
described by Eq. (\ref{ek7})) while we have no result for
$T_{c}<T<T_{*}$. In that range of temperatures we probably have a
homogeneous distribution of pairs.

\section{The case $N=2$}
\label{sec_n2}

The statistical mechanics of two particles in gravitational
interaction (binary star) was considered by Padmanabhan
\cite{paddy} in $d=3$ dimensions. He showed that this ``toy model''
exhibits phenomena that are representative of more realistic stellar
systems with a large number of particles. Here, we shall extend his
analysis in $d=2$ dimensions (the one-dimensional case is treated in
Appendix \ref{sec_ud}).  For $N=2$ particles, it is possible to
compute the density of states and the partition function exactly. This
will allow us to illustrate the preceding results on an explicit
example.

\subsection{The microcanonical ensemble}
\label{sec_dm}

The Hamiltonian of a system of two particles with mass $m_{1}$ and
$m_{2}$ in gravitational interaction can be written
\begin{eqnarray}
\label{dm1}
H=\frac{1}{2}MV^{2}+\frac{1}{2}\mu v^{2}+G m_{1}m_{2} \ln r,
\end{eqnarray}
where $({\bf R},{\bf V})$ denote the position and the velocity of the centre of mass and $({\bf r},{\bf v})$ the position and the velocity of the reduced particle. On the other hand
\begin{eqnarray}
\label{dm2}
M=m_{1}+m_{2}, \qquad \mu=\frac{m_{1}m_{2}}{m_{1}+m_{2}},
\end{eqnarray}
denote, respectively, the total mass of the particles and the mass of
the reduced particle.  We shall first compute the hypersurface of
phase space with energy less than $E$, i.e.
\begin{eqnarray}
\label{dm3}
\Gamma(E)=\int_{H\le E}d{\bf R}d{\bf V}d{\bf r}d{\bf v}.
\end{eqnarray}
Noting $x_{i}=(M/2)^{1/2}V_{i}$ for $i=1,2$ and  $x_{i}=(\mu/2)^{1/2}v_{i}$ for $i=3,4$, the preceding quantity can be rewritten 
\begin{eqnarray}
\label{dm4}
\Gamma(E)=\frac{4\pi R^{2}}{\mu M}\int d{\bf r}\int_{\| {\bf x}\|\le \sqrt{E-G m_{1}m_{2}\ln r}}d^{4}{x}.
\end{eqnarray}
The last integral represents the volume of a four-dimensional hypersphere with radius $\sqrt{E-G m_{1}m_{2}\ln r}$. Therefore, we obtain
\begin{eqnarray}
\label{dm5}
\Gamma(E)=\frac{8\pi^{2} R^{2}}{\mu M}V_{4}\int_{0}^{r_{m}} \left (E-G m_{1}m_{2}\ln r\right )^2 r dr,\nonumber\\
\end{eqnarray}
where $r_{m}$ denotes  the maximum distance accessible to the reduced particle and  $V_{4}$ is the volume of a four-dimensional
hypersphere with unit radius. From the general expression
\begin{eqnarray}
\label{dm5b}
V_{d}=\frac{\pi^{d/2}}{\Gamma(d/2+1)},
\end{eqnarray}
we find that $V_{4}=\pi^{2}/2$. The density of states in the microcanonical ensemble is given by
$g(E)=d\Gamma/dE$. Using Eq. (\ref{dm5}), we obtain
\begin{eqnarray}
\label{dm6}
g(E)=\frac{8\pi^{4} R^{2}}{\mu M}\int_{0}^{r_{m}} \left (E-G m_{1}m_{2}\ln r\right ) r dr.
\end{eqnarray}
The range of integration $r_{m}$ is such that $E-G m_{1}m_{2}\ln r\ge 0$ and $r\le R$. Therefore, $r_{m}=e^{E/Gm_{1}m_{2}}$ if $E\le G m_{1}m_{2}\ln R$ and 
 $r_{m}=R$ if $E\ge G m_{1}m_{2}\ln R$. Introducing
\begin{eqnarray}
\label{dm7}
A=8\pi^{4}G, \quad \epsilon=\frac{E}{Gm_{1}m_{2}}, \quad t=\frac{k_{B}T}{Gm_{1}m_{2}},
\end{eqnarray}
we obtain
\begin{eqnarray}
\label{dm8}
g(\epsilon)=AR^{2}\int_{0}^{r_{m}} \left (\epsilon-\ln r\right ) r dr,
\end{eqnarray}
with $r_{m}=e^{\epsilon}$ if $\epsilon\le \ln R$ and $r_{m}=R$ if $\epsilon\ge \ln R$. The entropy is given by
\begin{eqnarray}
\label{dm9}
S(\epsilon)=k_{B}\ln g(\epsilon),
\end{eqnarray}
and the temperature by
\begin{eqnarray}
\label{dm10}
\frac{1}{T}=\frac{\partial S}{\partial E}\quad \rightarrow \frac{1}{t}=\frac{\partial (S/k_{B})}{\partial \epsilon}.
\end{eqnarray}
Finally, the pressure is given by
\begin{eqnarray}
\label{dm11}
P=T\frac{\partial S}{\partial V}, \qquad V=\pi R^{2}.
\end{eqnarray}

\begin{figure}
\centering
\includegraphics[width=8cm]{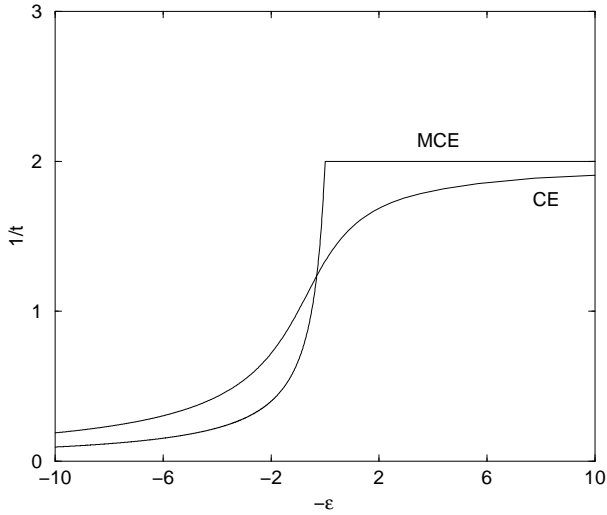}
\caption{Caloric curve in microcanonical (MCE) and canonical (CE) ensembles for the binary star model $N=2$ in two-dimensions. The caloric curves are similar in the two ensembles (without negative specific heat region) so that there is no phase transition contrary to the situation in $d=3$ \cite{paddy,rev}. There exists equilibrium states for all energies in MCE and for  temperatures $t>t_{*}=1/2$ in CE. It can be of interest to compare these curves  with the caloric curve obtained within the mean field approximation valid for $N\gg 1$ which presents a similar behaviour \cite{klb,sc}.    }
\label{caloricd2}
\end{figure}

The caloric curve $t=t(\epsilon)$ and the equation of state in the
microcanonical ensemble can be obtained from the exact expression
(\ref{dm8}) of the density of states. For $\epsilon\le \ln R$, we
obtain
\begin{eqnarray}
\label{dm12}
g(\epsilon)=\frac{AR^{2}}{4}e^{2\epsilon},
\end{eqnarray}
\begin{eqnarray}
\label{dm13}
k_{B}T=\frac{Gm_{1}m_{2}}{2}\quad \rightarrow t=\frac{1}{2},
\end{eqnarray}
\begin{eqnarray}
\label{dm14}
PV=k_{B}T.
\end{eqnarray}
For $\epsilon\ge \ln R$, we obtain
\begin{eqnarray}
\label{dm15}
g(\epsilon)=\frac{AR^{4}}{2}\left (\epsilon-\ln R+\frac{1}{2}\right ),
\end{eqnarray}
\begin{eqnarray}
\label{dm16}
t=\epsilon-\ln R+\frac{1}{2},
\end{eqnarray}
\begin{eqnarray}
\label{dm17}
PV=2k_{B}(T-T_{c}),
\end{eqnarray}
with 
\begin{eqnarray}
\label{dm18}
k_{B}T_{c}=\frac{G m_{1}m_{2}}{4}.
\end{eqnarray}
The caloric curve $t(\epsilon)$ in the microcanonical ensemble is represented in Fig. \ref{caloricd2}.

\subsection{The canonical ensemble}
\label{sec_ddc}

\begin{figure}
\centering
\includegraphics[width=8cm]{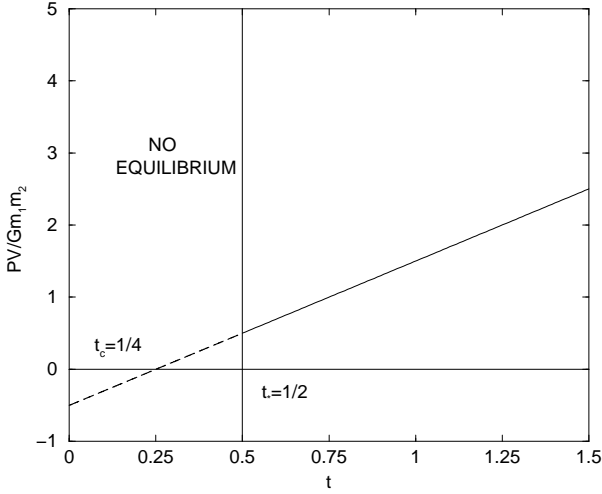}
\caption{Equation of state for the binary star model $N=2$ in two-dimensions. Equilibrium states exist only for $t>t_{*}=1/2$ so that the portion of the curve $t_{c}=1/4<t<t_{*}=1/2$ is not physical. Note that the pressure does not vanish at $t=t_{*}=1/2$. It corresponds to the pressure due to the Dirac peak (one effective particle) that forms at this critical temperature (see Sec. \ref{sec_tt}). }
\label{eosd2}
\end{figure}

For $N=2$ particles in gravitational interaction, the partition
function in the canonical ensemble is given by
\begin{eqnarray}
\label{ddc1}
Z=\int e^{-\beta M\frac{V^{2}}{2}}e^{-\beta \mu\frac{v^{2}}{2}}
e^{-\beta m_{1}m_{2}G\ln r}d{\bf R}d{\bf V}d{\bf r}d{\bf v}.\nonumber\\
\end{eqnarray}
Integrating on ${\bf R}$  and on the velocities, we get 
\begin{eqnarray}
\label{ddc2}
Z=\frac{8\pi^{3}V}{\beta^{2}m_{1}m_{2}}\int_{0}^{R}
r^{1-\beta G m_{1}m_{2}}\ dr.
\end{eqnarray}
The partition function is finite if, and only if, $T>T_{*}$ with
\begin{eqnarray}
\label{ddc3}
k_{B}T_{*}=\frac{Gm_{1}m_{2}}{2}.
\end{eqnarray}
For equal mass particles, we have
\begin{eqnarray}
\label{ddc4}
k_{B}T_{*}=\frac{Gm^{2}}{2}.
\end{eqnarray}
These expressions agree with the general results of
Sec. \ref{sec_existence}. For $N=2$, we have $T_{1}=T_{2}=T_{*}$ even
if the particles have different masses $m_{1}\neq m_{2}$. For
$T>T_{*}$, the partition function is explicitly given by
\begin{eqnarray}
\label{ddc5}
Z=\frac{8\pi^{4}}{\beta^{2}m_{1}m_{2}}\frac{R^{4-\beta Gm_{1}m_{2}}}{2-\beta Gm_{1}m_{2}}.
\end{eqnarray}
This expression corresponds to the bound obtained in Eq. (\ref{ex8}).
The pressure can be computed from Eqs. (\ref{ca5}) and (\ref{ddc5}) leading to
\begin{eqnarray}
\label{ddc6}
PV=2k_{B}(T-T_{c}),
\end{eqnarray}
with
\begin{eqnarray}
\label{ddc7}
k_{B}T_{c}=\frac{Gm_{1}m_{2}}{4}.
\end{eqnarray}
For equal mass particles, we get
\begin{eqnarray}
\label{ddc8}
k_{B}T_{c}=\frac{Gm^{2}}{4}.
\end{eqnarray}
These expressions are consistent with Eqs. (\ref{ca7})-(\ref{ca9}). We also check
explicitly that $T_{*}\neq T_{c}$ (for $N=2$, we have
$T_{*}=2T_{c}$). Therefore, $T_{c}$ is smaller than the collapse
temperature $T_{*}$ below which the partition function diverges.
The curve $P(T)$ is plotted in Fig. \ref{eosd2}.

\begin{figure}
\centering
\includegraphics[width=8cm]{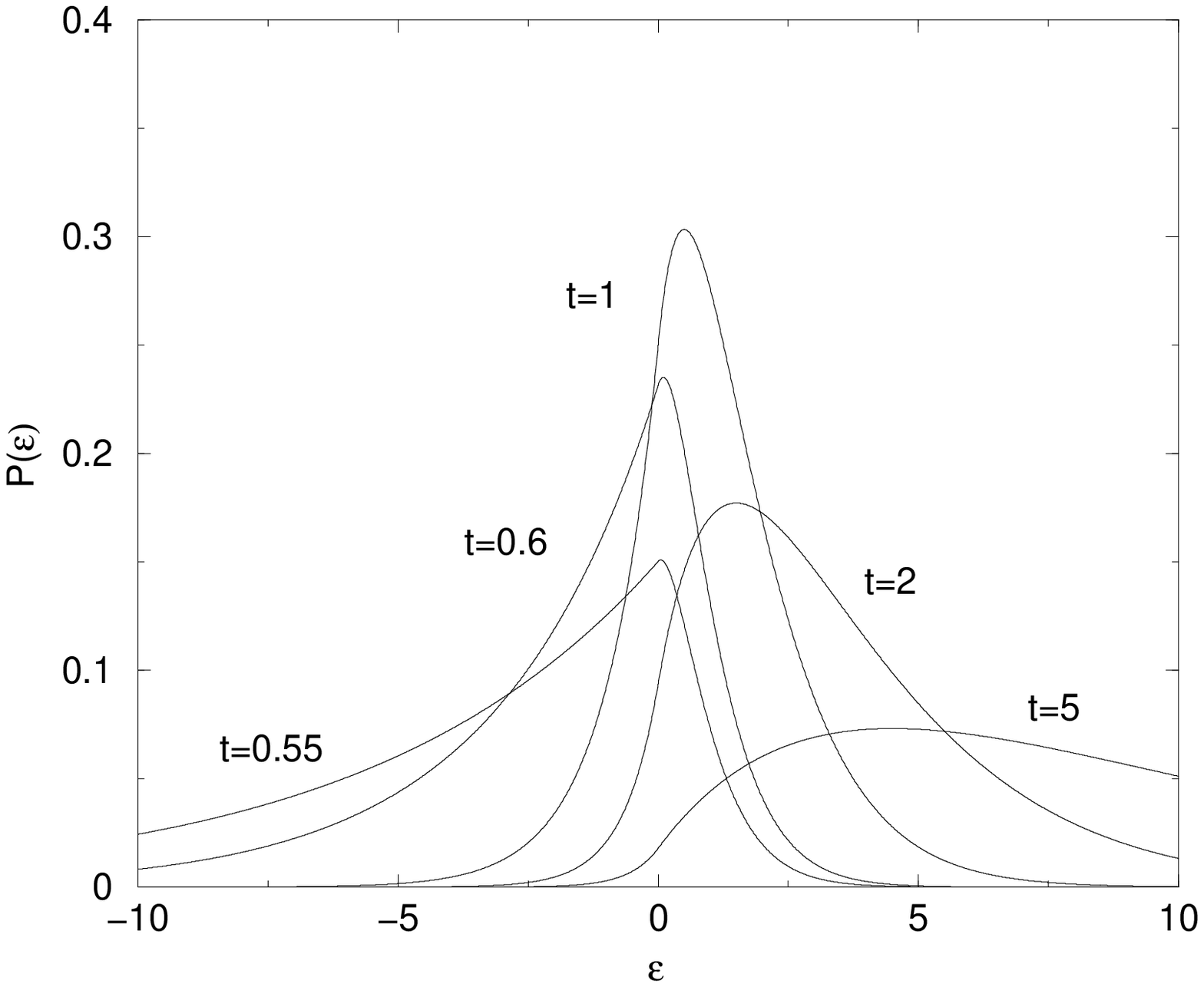}
\caption{Distribution of energies for the binary star model $N=2$ in two-dimensions in the canonical ensemble. We have plotted this distribution for different values of the temperature. }
\label{prob2}
\end{figure}

The caloric curve $\langle \epsilon\rangle (t)$ in the canonical
ensemble can be obtained from the exact expression (\ref{ddc5}) of the
partition function. The average energy is given by
\begin{eqnarray}
\label{ddc9}
\langle E\rangle =-\frac{\partial \ln Z}{\partial\beta} \quad \rightarrow \langle\epsilon\rangle =t^{2}\frac{\partial\ln Z}{\partial t}.
\end{eqnarray}
Using Eq. (\ref{ddc5}), we obtain
\begin{eqnarray}
\label{ddc10}
\langle\epsilon\rangle=\frac{(4t-3)t}{2t-1}+\ln R.
\end{eqnarray}
The caloric curve $\langle\epsilon\rangle(t)$ in the canonical
ensemble is represented in Fig. \ref{caloricd2} and is compared with
the caloric curve in the microcanonical ensemble.  We note that, for
$N=2$, the specific heats $C=d\langle E\rangle/dT$ diverges like $\propto
(T-T_{*})^{-2}$ at $T=T_{*}$. By contrast, in the limit $N\rightarrow
+\infty$ where a mean field approximation can be implemented, we find
from Eq. (40) of Sire \& Chavanis \cite{sc} that $C\propto
(T-T_{*})^{-1}$.

Finally, the distribution of energies at temperature $T$ in the
canonical ensemble is given by
\begin{eqnarray}
\label{ddc11}
P(E)=\frac{1}{Z(T)}g(E)e^{-\beta E}.
\end{eqnarray}
Using $P(E)dE=P(\epsilon)d\epsilon$, this can be rewritten
\begin{eqnarray}
\label{ddc12}
\frac{P(\epsilon)}{Gm_{1}m_{2}}=\frac{1}{Z(t)}g(\epsilon)e^{-\epsilon/t}.
\end{eqnarray}
Using Eqs. (\ref{dm12}), (\ref{dm15}) and (\ref{ddc5}), we find that
\begin{eqnarray}
\label{ddc13}
P(\epsilon)=\frac{2t-1}{4t^{3}}e^{(2-1/t)\epsilon}, \qquad (\epsilon\le 0),
\end{eqnarray}
\begin{eqnarray}
\label{ddc14}
P(\epsilon)=\frac{2t-1}{4t^{3}}(2\epsilon+1)e^{-\epsilon/t}, \qquad (\epsilon\ge 0).
\end{eqnarray}
We have taken $R=1$ to simplify the expressions. The distribution of energies is represented in Fig. \ref{prob2} for different temperatures.

Taking $G=-1$ and $m_{i}=q_{i}$, these results also describe a
two-dimensional plasma with $1$ charge $+e$ and $1$ charge
$-e$. According to Eq. (\ref{ddc3}), the partition function exists
only for $T>T_{*}$ with
\begin{eqnarray}
\label{ddc15ne}
k_{B}T_{*}={e^{2}\over 2},
\end{eqnarray}   
while the critical temperature  appearing in the equation of state (\ref{ddc6}) is given by 
\begin{eqnarray}
\label{ddc15nebtr}
k_{B}T_{c}={e^{2}\over 4}.
\end{eqnarray} 
Thus
\begin{eqnarray}
\label{ddc15bbis}
T_{*}=2T_{c}.
\end{eqnarray}  
More general studies \cite{alastuey} show that the expression
(\ref{ddc15ne}) of the collapse temperature remains valid for $N/2$
charges $+e$ and $N/2$ charges $-e$ where $N$ is arbitrary. Indeed,
the ``catastrophic collapse'' at $T=T_{*}$ corresponds to the
formation of $N_{*}=N/2$ non-interacting pairs $(+,-)$ such as those
studied here {\it individually}. It can therefore be obtained by
studying the collapse of just {\it one} pair. Therefore, in 2D plasma
physics, $T_{*}=2T_{c}$ for any $N$ (for a neutral plasma with two
components).  By comparison, in 2D gravity $T_{*}={N\over N-1}T_{c}$
(for a one component gas).

\subsection{Relevance of a statistical description for $N=2$ particles}
\label{sec_rel}

We may wonder about the relevance of studying a system of $N=2$
particles using statistical ensembles.

For Brownian particles, the $N$-body distribution function is governed
by the Fokker-Planck equation (\ref{ek3}) whose steady state is the
canonical distribution (\ref{ek5}). This is valid for {\it any}
$N$. For $N=1$, we recover the ordinary Brownian motion of a free
particle submitted to a friction and a noise. The equilibrium state is
the Maxwell-Boltzmann distribution predicted by statistical
mechanics. For $N=2$, we have two Brownian particles interacting via
self-gravity and this system can be studied using the canonical
ensemble. More generally, for Brownian particles, the canonical ensemble is
justified for any $N$. For Hamiltonian systems, the relevance of a
statistical description for a small number of particles is less
clear. For example, $N=2$ particles in gravitational interaction have
a simple deterministic motion satisfying the Kepler laws. A
statistical description may be justified, however, when the particles
are hard spheres enclosed within a box so that their motion is very
irregular due to collisions between themselves or against the walls
(see the discussion in the review of Padmanabhan
\cite{paddy}). In that case, the microcanonical ensemble may be employed.

In any case, the density of states and the partition function can be
defined mathematically for any $N$ and it is interesting to compute
these quantities exactly for $N=2$ and compare these results with the
ones obtained for $N\rightarrow +\infty$ by using a mean field
approximation.  For 3D self-gravitating systems, it is shown in
\cite{paddy,rev} that the caloric curves (in microcanonical and
canonical ensembles) obtained with $N=2$ particles are already very
close to those obtained for a large number of particles. Similarly, we
have shown in the present paper that the same observation holds in
$d=2$ and $d=1$ dimensions. This means that the mean field results are
 reliable even for a small number of particles.

\section{Conclusion: analogies with chemotaxis and point vortex dynamics}
\label{sec_conclusion}

In this paper, we have obtained the exact expression of the diffusion
coefficient of a gas of self-gravitating Brownian particles in two
dimensions. Two-dimensional gravity is a rare example where exact
results can be obtained \footnote{Of course, the results of this paper
can be easily extended to power-law potentials of the form
$u'(\xi)=G/\xi^{d-\alpha}$ in $d$ dimensions. In that case, the
critical dimension is $d_{crit}=\alpha+1$.}. The diffusion coefficient
exhibits a critical temperature $T_{c}$ below which it becomes
negative indicating finite time blow-up. In a finite domain, this
critical temperature is slightly smaller than the collapse temperature
$T_{*}$ below which the partition function diverges. However, this
difference is small, the ratio being of the order of $1-1/N$, and the
two temperatures coincide at the thermodynamic (mean field) limit
$N\rightarrow +\infty$. We may wonder what happens dynamically in the
region $T_{c}<T<T_{*}$ since there is no equilibrium state in that
case while we cannot prove finite time blow up from the Virial
theorem. The system collapses for $T<T_{*}$ but it is possible that,
for $T_{c}<T<T_{*}$, it takes an infinite time to form the Dirac
peak. We stress that the computation of the diffusion coefficient
brings only partial information on the dynamics of the
self-gravitating Brownian gas. For example, it does not give any
information concerning the precise evolution of the density
distribution of the particles. To obtain the dynamical evolution of
the density profile $\rho({\bf r},t)$, we need to solve the $N$-body
equations (\ref{ek1})-(\ref{ek2}) or make a mean field approximation,
valid in the proper thermodynamic limit $N\rightarrow +\infty$, where
the problem reduces to the study of the Smoluchowski-Poisson (SP)
system. Therefore, the complete description of the self-gravitating
Brownian gas remains highly complicated and requires in general some
approximations even if an exact result (\ref{intro1})-(\ref{intro2})
can be obtained.

The results obtained in this paper can also be relevant in
mathematical biology \cite{murray}. They apply to a simplified version of the
Keller-Segel model \cite{ks} modelling the chemotaxis of bacterial populations.  The Keller-Segel model can be derived from a system of
coupled Langevin equations of the form \cite{newman,kinbio}:
\begin{eqnarray}
\label{conc1}
\frac{d{\bf r}_{\alpha}}{dt}=\chi\nabla_{\alpha}c+\sqrt{2D}{\bf R}_{\alpha}(t),
\end{eqnarray}
\begin{eqnarray}
\label{conc2}
\frac{\partial c}{\partial t}=-kc+D_{c}\Delta c+h\sum_{\alpha=1}^{N}\delta({\bf r}-{\bf r}_{\alpha}(t)).
\end{eqnarray}
Here, ${\bf r}_{\alpha}(t)$ represents the position of a biological
entity (cell, amoeba, bacteria,...) and $c({\bf r},t)$ is the
concentration of the secreted chemical. The cells have a diffusion
motion with diffusion coefficient $D$ and they also move along a
positive gradient of chemical (attractive chemotaxis $\chi>0$). The
chemical is produced by the cells with a strength $h$. It is in
addition degraded at a rate $k$ and diffuses with a diffusion
coefficient $D_{c}$. The usual Keller-Segel model is recovered  from
these equations by making a mean field approximation leading to
\begin{eqnarray}
\label{conc3}
\frac{\partial \rho}{\partial t}=D\Delta\rho-\chi\nabla\cdot (\rho\nabla c),
\end{eqnarray}
\begin{eqnarray}
\label{conc4}
\frac{\partial c}{\partial t}=-kc+D_{c}\Delta c+h\rho.
\end{eqnarray}
Some authors have considered a simplified chemotactic model where the
equation (\ref{conc2}) for the chemical is replaced by a Poisson
equation
\begin{eqnarray}
\label{conc5}
\Delta c=-\lambda \sum_{\alpha=1}^{N} m\delta({\bf r}-{\bf r}_{\alpha}(t)).
\end{eqnarray}
This can be justified in a limit of high diffusivity of the chemical
and, in addition, for sufficiently dense systems
\cite{jager,kinbio}. In that case, the chemotactic problem becomes
isomorphic to the study of self-gravitating Brownian particles
described by Eq. (\ref{ek7}), with a proper re-interpretation of the
parameters. In particular, the concentration $-c$ of the chemical
plays the role of the gravitational potential $\Phi$. We have the
additional correspondances $\lambda\leftrightarrow S_{d}G$ and
$\chi\leftrightarrow \mu m$. All the results of this paper can then be
extended to the biological context provided that we make the
appropriate substitutions. Furthermore, the consideration of the
dimension $d=2$ is particularly justified in biology since cellular
organisms are often ascribed to move on a plane. In chemotaxis, we
rather use the mass as a control parameter instead of the
temperature. For example, the critical mass corresponding to the
critical temperature $T_{c}$ is
\begin{eqnarray}
\label{conc6}
M_{c}=\frac{8\pi D}{\lambda\chi}+m,
\end{eqnarray} 
and the collapse mass  corresponding to the collapse temperature $T_{*}$ is
\begin{eqnarray}
\label{conc7}
M_{*}=\frac{8\pi D}{\lambda\chi}.
\end{eqnarray}
For a large number of cells ($M\gg m$) the two values coincide. For
$M<M_{*}$, a box-confined system reaches an equilibrium state.  For
$M>M_{*}$, the system collapses and forms a Dirac peak corresponding
to chemotactic aggregation. The adaptation of our results to the
biological context is straightforward  \cite{masstemp}  so it will not
be discussed in more detail here.

Let us finally consider the case of point vortices in two-dimensional
hydrodynamics (see, e.g., \cite{houches}). The Hamiltonian of the
point vortex gas is \footnote{For brevity, we do not take into account
the contribution of the images (in a bounded domain) because most of
the following results are obtained by considering configurations where
the vortices form compact clusters. They are therefore independent on
boundary effects.}
\begin{eqnarray}
\label{conc8}
H=-\frac{1}{2\pi}\sum_{i<j}\gamma_{i}\gamma_{j}\ln |{\bf r}_{i}-{\bf r}_{j}|,
\end{eqnarray}
where the coordinates $(x,y)$ are canonically conjugate
\cite{kirchhoff}.  The partition function in the canonical ensemble 
 is given by
\begin{eqnarray}
\label{conc9}
Z=\int e^{\frac{\beta}{2\pi}\sum_{i<j}\gamma_{i}\gamma_{j}\ln |{\bf
r}_{i}-{\bf r}_{j}|}\prod_{k=1}^{N} d{\bf r}_{k}.
\end{eqnarray}
We note that there is no kinetic (quadratic) term in the Hamiltonian
(\ref{conc8}) in the usual sense, so that the temperature can take
both positive or {\it negative} values \cite{onsager}. By contrast,
only positive temperatures are allowed in plasma physics and
gravity. Let us first assume that the point vortices have the same
circulation $\gamma$.  At negative temperatures, the partition
function is formally equivalent to that for self-gravitating
systems. Therefore, at very negative inverse temperatures, point
vortices have the tendency to ``attract'' each other and group
themselves in a single aggregate (supervortex) of circulation
$N\gamma$. This corresponds to a regime of high energies. This is
similar to the formation of a Dirac peak in the gravitational problem
for $T<T_{*}$. Thus, the partition function exists if and only if
$\beta>\beta_{*}$ with
\begin{eqnarray}
\label{conc10}
\beta_{*}=-\frac{8\pi}{N\gamma^{2}}.
\end{eqnarray}
This is the exact equivalent of the collapse temperature (\ref{tt1})
as can be seen by taking $G=-1/(2\pi)$ and $m\leftrightarrow
\gamma$. At positive temperatures,  the partition
function is formally equivalent to that for a non-neutral plasma of
equal charges. Therefore, at large positive inverse temperatures,
point vortices have the tendency to ``repell'' each other and
accumulate on the boundary of the domain.  We now consider $N$
vortices with circulations $+\gamma$ and $N$ vortices with
circulations $-\gamma$. At positive temperatures, the partition
function is formally equivalent to that for a neutral
plasma. Therefore, at large positive inverse temperatures, the
vortices of opposite sign have the tendency to ``attract'' each other
and form $N$ dipoles $(+,-)$. This corresponds to a regime of high
negative energies. This is similar to the formation of $N$ pairs
$(+,-)$ in plasma physics for $T<T_{*}$.  Thus, the partition function
exists if and only if $\beta<\beta_{*}$ with
\begin{eqnarray}
\label{conc11}
\beta_{*}=\frac{4\pi}{\gamma^{2}}.
\end{eqnarray}
This is the exact equivalent of the collapse temperature
(\ref{ddc15}).  At negative temperatures, the vortices of same sign
have the tendency to ``attract'' each other and form an aggregate of
positive circulation $N\gamma$ and an aggregate of negative
circulation $-N\gamma$ (vortex dipole). This corresponds to a regime
of high positive energies. This is similar to the formation of a Dirac
peak in the gravitational problem, except that we have here {\it two}
peaks: one with positive circulation and one with negative
circulation. The partition function exists if and only if
$\beta>\beta_{*}$ where $\beta_{*}$ is given by Eq. (\ref{conc10})
like in the case with one cluster of $N$ particles. Therefore, the
point vortex gas is very rich because it exhibits features similar to
self-gravitating systems at negative temperatures and features similar
to plasma systems at positive temperatures. We emphasize, however,
that point vortices form a Hamiltonian system described by the
microcanonical ensemble so that the control parameter is the energy
(not the temperature) and the quantity of interest is the density of
states (not the partition function). It can be shown that the density
of states converges for any accessible energy. Therefore, the above
collapse temperatures (\ref{conc10})-(\ref{conc11}) represent lower
and upper bounds on the caloric curve $\beta(E)$ which correspond to
$E\rightarrow +\infty$ and $E\rightarrow -\infty$ respectively. The
question to know what happens for $\beta<-\frac{8\pi}{N\gamma^{2}}$ or
$\beta>\frac{4\pi}{\gamma^{2}}$ has {\it a priori} no sense in vortex
dynamics since we cannot impose the temperature of the vortex gas.

\vskip0.5cm
{\it Acknowledgements} I acknowledge stimulating discussions with
C. Sire and D. Dean on this subject. I am also grateful to D. Dalli\'e
for encouragements.

\appendix

\section{Alternative derivation of the kinetic equation}
\label{sec_a}

In this Appendix, we provide an alternative derivation of the exact
kinetic equation Eq. (\ref{ek16}). The Langevin equation (\ref{ek7})
governing the time evolution of particle $\alpha$ in the strong
friction limit can be written
\begin{eqnarray}
\label{a1}
\frac{d{\bf r}_{\alpha}}{dt}=-m_{\alpha}\mu_{\alpha}\nabla\Phi^{(\alpha)}({\bf r}_{\alpha})+\sqrt{2D_{\alpha}}{\bf R}_{\alpha}(t),
\end{eqnarray}
where
\begin{eqnarray}
\label{a2}
\Phi^{(\alpha)}({\bf r},t)=\sum_{\beta\neq\alpha} m_{\beta}u(|{\bf r}-{\bf r}_{\beta}(t)|),
\end{eqnarray}
denotes the exact potential created by the other particles.  We define
the one and two-body probability distributions  by
\begin{eqnarray}
\label{a3}
P_{\alpha}({\bf r},t)=\langle \delta({\bf r}-{\bf r}_{\alpha}(t))\rangle,
\end{eqnarray}
\begin{eqnarray}
\label{a4}
P_{\alpha,\beta}({\bf r},{\bf r}',t)=\langle \delta({\bf r}-{\bf r}_{\alpha}(t))\delta({\bf r}'-{\bf r}_{\beta}(t))\rangle.
\end{eqnarray}
Differentiating Eq. (\ref{a3}) with respect to time, we obtain
\begin{eqnarray}
\label{a5}
\frac{\partial P_{\alpha}}{\partial t}=-\nabla\cdot \langle {\dot {\bf r}}_{\alpha}(t)\delta({\bf r}-{\bf r}_{\alpha}(t))\rangle.
\end{eqnarray}
Substituting from Eq. (\ref{a1}) in Eq. (\ref{a5}), we find that
\begin{eqnarray}
\label{a6}
\frac{\partial P_{\alpha}}{\partial t}=-\nabla\cdot \langle \sqrt{2D_{\alpha}}{\bf R}_{\alpha}(t) \delta({\bf r}-{\bf r}_{\alpha}(t))\rangle\nonumber\\
 +\nabla\cdot \langle m_{\alpha}\mu_{\alpha}\nabla\Phi^{(\alpha)}({\bf r}_{\alpha}(t))\delta({\bf r}-{\bf r}_{\alpha}(t))\rangle .
\end{eqnarray}
The first term is the standard term that appears in deriving the Fokker-Planck equation for a pure random walk; it leads to a term proportional to the Laplacian of $P_{\alpha}$:
\begin{eqnarray}
\label{a6bis}
-\nabla\cdot \langle \sqrt{2D_{\alpha}}{\bf R}_{\alpha}(t) \delta({\bf r}-{\bf r}_{\alpha}(t))\rangle=D_{\alpha}\Delta P_{\alpha}.\nonumber\\
\end{eqnarray}
To evaluate the  second term, we first write
\begin{eqnarray}
\label{a7}
\Phi^{(\alpha)}({\bf r},t)=\int d{\bf r}' \sum_{\beta\neq\alpha} m_{\beta}u(|{\bf r}-{\bf r}'|)\delta({\bf r}'-{\bf r}_{\beta}(t)).\nonumber\\
\end{eqnarray}
This yields
\begin{eqnarray}
\label{a8}
\langle \nabla\Phi^{(\alpha)}({\bf r}_{\alpha}(t))\delta({\bf r}-{\bf r}_{\alpha}(t))\rangle= \nonumber\\
\int d{\bf r}' \sum_{\beta\neq\alpha} m_{\beta} \nabla u(|{\bf r}-{\bf r}'|)\langle \delta({\bf r}-{\bf r}_{\alpha}(t)) \delta({\bf r}'-{\bf r}_{\beta}(t))\rangle\nonumber\\
=\int d{\bf r}' \sum_{\beta\neq\alpha} m_{\beta} P_{\alpha,\beta}({\bf r},{\bf r}',t) \nabla u(|{\bf r}-{\bf r}'|).\qquad
\end{eqnarray}
Substituting Eqs. (\ref{a6bis}) and (\ref{a8}) in Eq. (\ref{a6}), we get the exact kinetic equation (\ref{ek16}).

We can easily extend this approach so as to take into account the
inertia of the particles. In that case, the dynamical evolution of
particle $\alpha$ is described by stochastic equations of the form
\begin{equation}
\label{a9}{d{\bf r}_{\alpha}\over dt}={\bf v}_{\alpha},
\end{equation}
\begin{equation}
\label{a10}
 {d{\bf v}_{\alpha}\over dt}=-\xi_{\alpha} {\bf v}_{\alpha}-\nabla\Phi^{(\alpha)}({\bf r}_{\alpha}(t))+\sqrt{2D'_{\alpha}}{\bf R}_{\alpha}(t).
\end{equation}
We define the one-body probability distributions in phase space by
\begin{equation}
\label{a11}P_{\alpha}({\bf r},{\bf v},t)=\langle \delta({\bf r}-{\bf r}_{\alpha}(t))\delta({\bf v}-{\bf v}_{\alpha}(t))\rangle,
\end{equation}
where the brackets denote an average over the noise. Similarly, the  two-body probability distribution in phase space is
\begin{eqnarray}
\label{a12}P_{\alpha,\beta}({\bf r},{\bf v};{\bf r}',{\bf v}',t)=\langle \delta({\bf r}-{\bf r}_{\alpha}(t))\delta({\bf v}-{\bf v}_{\alpha}(t))\nonumber\\
\delta({\bf r}'-{\bf r}_{\beta}(t))\delta({\bf v}'-{\bf v}_{\beta}(t))\rangle.
\end{eqnarray} 
Taking the time derivative of $P_{\alpha}$, we get
\begin{eqnarray}
\label{a13}{\partial P_{\alpha}\over\partial t}=-{\partial\over\partial {\bf r}}\cdot \langle {\dot{\bf r}}_{\alpha}(t) \delta({\bf r}-{\bf r}_{\alpha}(t))\delta({\bf v}-{\bf v}_{\alpha}(t))\rangle\nonumber\\
 -{\partial\over\partial {\bf v}}\cdot \langle {\dot{\bf v}}_{\alpha}(t) \delta({\bf r}-{\bf r}_{\alpha}(t))\delta({\bf v}-{\bf v}_{\alpha}(t))\rangle . 
\end{eqnarray} 
Inserting the equations of motion (\ref{a9})-(\ref{a10}) in
Eq. (\ref{a13}), we obtain
\begin{eqnarray}
\label{a14}{\partial P_{\alpha}\over\partial t}=-{\partial\over\partial {\bf r}}\cdot \langle {{\bf v}}_{\alpha}(t)\ \delta({\bf r}-{\bf r}_{\alpha}(t))\delta({\bf v}-{\bf v}_{\alpha}(t))\rangle\nonumber\\
 +{\partial\over\partial {\bf v}}\cdot\langle \xi_{\alpha} {{\bf v}}_{\alpha}(t)\ \delta({\bf r}-{\bf r}_{\alpha}(t))\delta({\bf v}-{\bf v}_{\alpha}(t))\rangle \nonumber\\ 
+{\partial\over\partial {\bf v}}\cdot \langle \nabla\Phi^{(\alpha)}({\bf r}_{\alpha}(t)) \ \delta({\bf r}-{\bf r}_{\alpha}(t))\delta({\bf v}-{\bf v}_{\alpha}(t))\rangle\nonumber\\
-{\partial\over\partial {\bf v}}\cdot\langle \sqrt{2D'_{\alpha}} {\bf R}_{\alpha}(t) \ \delta({\bf r}-{\bf r}_{\alpha}(t))\delta({\bf v}-{\bf v}_{\alpha}(t))\rangle.  \end{eqnarray} 
The first two terms are straightforward to evaluate. The fourth term is the standard term that appears in deriving the Fokker-Planck equation for a pure random walk; it leads to a term proportional to the Laplacian of $P_{\alpha}$ in velocity space.  The third term can be evaluated by inserting the expression (\ref{a7}) in Eq. (\ref{a14}). Finally, we obtain the exact equation
\begin{eqnarray}
\label{a15}{\partial P_{\alpha}\over\partial t}+{\bf v}\cdot {\partial P_{\alpha}\over\partial {\bf r}}-{\partial\over\partial {\bf v}}\cdot \int d{\bf r}'\sum_{\beta\neq\alpha} m_{\beta} \nabla u (|{\bf r}-{\bf r}'|)\nonumber\\
\times P_{\alpha,\beta}({\bf r},{\bf v};{\bf r}',t)
={\partial\over\partial {\bf v}}\cdot \left \lbrack D'_{\alpha}{\partial P_{\alpha}\over\partial {\bf v}}+\xi_{\alpha} P_{\alpha}{\bf v}\right\rbrack,  \qquad\qquad  
\end{eqnarray} 
where the statistical correlations are encapsulated in the two-body
probability distribution
\begin{eqnarray}
\label{a16}P_{\alpha,\beta}({\bf r},{\bf v};{\bf r}',t)=\langle \delta({\bf r}-{\bf r}_{\alpha}(t))\delta({\bf v}-{\bf v}_{\alpha}(t))
\delta({\bf r}'-{\bf r}_{\beta}(t))\rangle.\nonumber\\
\end{eqnarray} 
If we implement a mean field approximation
\begin{eqnarray}
\label{a17}P_{\alpha,\beta}({\bf r},{\bf v};{\bf r}',t)=P_{\alpha}({\bf r},{\bf v},t)P_{\beta}({\bf r}',t),
\end{eqnarray} 
the preceding equation can be rewritten
\begin{eqnarray}
\label{a18}{\partial P_{\alpha}\over\partial t}+{\bf v}\cdot {\partial P_{\alpha}\over\partial {\bf r}}-
 {\partial P_{\alpha}\over\partial {\bf v}}\cdot \nabla\int d{\bf r}'\sum_{\beta\neq\alpha} m_{\beta}
 u (|{\bf r}-{\bf r}'|)P_{\beta}({\bf r}',t)
\nonumber\\
={\partial\over\partial {\bf v}}\cdot \left \lbrack D'_{\alpha}{\partial P_{\alpha}\over\partial {\bf v}}+\xi_{\alpha} P_{\alpha}{\bf v}\right\rbrack.  \qquad\qquad  
\end{eqnarray} 
If we extend the sum over all $\beta$ and introduce the spatial density (\ref{mf3}), we obtain
\begin{eqnarray}
\label{a19}{\partial P_{\alpha}\over\partial t}+{\bf v}\cdot {\partial P_{\alpha}\over\partial {\bf r}}-{\partial P_{\alpha}\over\partial {\bf v}}\cdot
\nabla \int d{\bf r}'
 u (|{\bf r}-{\bf r}'|)\rho({\bf r}',t)
 \nonumber\\
={\partial\over\partial {\bf v}}\cdot \left \lbrack D'_{\alpha}{\partial P_{\alpha}\over\partial {\bf v}}+\xi_{\alpha} P_{\alpha}{\bf v}\right\rbrack,  \qquad\qquad  
\end{eqnarray} 
This can be rewritten in the form of a mean field Kramers equation
\begin{eqnarray}
\label{a20}{\partial P_{\alpha}\over\partial t}+{\bf v}\cdot {\partial P_{\alpha}\over\partial {\bf r}}-\nabla\Phi\cdot {\partial P_{\alpha}\over\partial {\bf v}}
={\partial\over\partial {\bf v}}\cdot \left \lbrack D'_{\alpha}{\partial P_{\alpha}\over\partial {\bf v}}+\xi_{\alpha} P_{\alpha}{\bf v}\right\rbrack, \nonumber\\ 
\end{eqnarray} 
where 
\begin{eqnarray}
\label{a21}
\Phi({\bf r},t)=\int
 \rho({\bf r}',t) u (|{\bf r}-{\bf r}'|) d{\bf r}',
\end{eqnarray}  
is the potential.

\section{The case $T=0$}
\label{sec_tz}

Here, we assume that all the particles have the same mass $m$. At $T=0$ (no noise), the equations of motion (\ref{ek7}) in the overdamped limit become 
\begin{eqnarray}
\label{tz1}
\frac{d{\bf r}_{\alpha}}{dt}=-\frac{1}{\xi}\nabla\Phi_{ex}({\bf r}_{\alpha},t),
\end{eqnarray}
where $\Phi_{ex}({\bf r},t)$ is the exact gravitational potential that is solution of the Poisson equation
\begin{eqnarray}
\label{tz2}
\Delta\Phi_{ex}=S_{d}G\rho_{ex},
\end{eqnarray}
with the exact density field
\begin{eqnarray}
\label{tz3}
\rho_{ex}({\bf r},t)=\sum_{\alpha=1}^{N}m\delta({\bf r}-{\bf r}_{\alpha}(t)).
\end{eqnarray}
Taking the time derivative of Eq. (\ref{tz3}), we get
\begin{eqnarray}
\label{tz4}
\frac{\partial \rho_{ex}}{\partial t}=-\sum_{\alpha=1}^{N}m \nabla \cdot  \left ({\dot {\bf r}}_{\alpha} \delta({\bf r}-{\bf r}_{\alpha}(t))\right ).
\end{eqnarray}
Substituting Eq. (\ref{tz1}) in Eq. (\ref{tz4}), we obtain
\begin{eqnarray}
\label{tz5}
\frac{\partial \rho_{ex}}{\partial t}=\frac{1}{\xi}\sum_{\alpha=1}^{N}m \nabla \cdot  \left (\nabla\Phi_{ex}({\bf r}_{\alpha}(t),t) \delta({\bf r}-{\bf r}_{\alpha}(t))\right )\nonumber\\
=\frac{1}{\xi}\nabla \cdot  \left (\nabla\Phi_{ex}({\bf r},t)\sum_{\alpha=1}^{N}m  \delta({\bf r}-{\bf r}_{\alpha}(t))\right ),
\end{eqnarray}
so that, finally,
\begin{eqnarray}
\label{tz6}
\frac{\partial \rho_{ex}}{\partial t}=\frac{1}{\xi}\nabla \cdot  \left (\rho_{ex}\nabla\Phi_{ex}({\bf r},t)\right ).
\end{eqnarray}
This exact equation, where $\rho_{ex}$ is expressed in terms of Dirac
functions and $\Phi_{ex}$ is given by Eq. (\ref{tz2}), contains
exactly the same information as the deterministic equations
(\ref{tz1}).

For the inertial model, at $T=0$, the equations of 
motion are
\begin{eqnarray}
\label{tz7}
\frac{d{\bf r}_{\alpha}}{dt}={\bf v}_{\alpha},
\end{eqnarray}
\begin{eqnarray}
\label{tz8}
\frac{d{\bf v}_{\alpha}}{dt}=-\xi {\bf v}_{\alpha}-\nabla\Phi_{ex}({\bf r}_{\alpha},t).
\end{eqnarray} 
Introducing the exact distribution function
\begin{eqnarray}
\label{tz9}
f_{ex}({\bf r},{\bf v},t)=\sum_{\alpha=1}^{N}m\delta({\bf r}-{\bf r}_{\alpha}(t))\delta({\bf v}-{\bf v}_{\alpha}(t)),
\end{eqnarray}
and taking the time derivative of Eq. (\ref{tz9}), we get
\begin{eqnarray}
\label{tz10}
\frac{\partial f_{ex}}{\partial t}=-\sum_{\alpha=1}^{N}m \nabla \cdot  \left ({\dot {\bf r}}_{\alpha} \delta({\bf r}-{\bf r}_{\alpha}(t))\delta({\bf v}-{\bf v}_{\alpha}(t))\right )\nonumber\\
-\sum_{\alpha=1}^{N}m \frac{\partial}{\partial {\bf v}} \cdot  \left ({\dot {\bf v}}_{\alpha} \delta({\bf r}-{\bf r}_{\alpha}(t))\delta({\bf v}-{\bf v}_{\alpha}(t))\right ).\nonumber\\
\end{eqnarray}
Substituting Eqs. (\ref{tz7})-(\ref{tz8}) in Eq. (\ref{tz10}) and following a procedure similar to that developed previously, we obtain
\begin{eqnarray}
\label{tz11}
\frac{\partial f_{ex}}{\partial t}+{\bf v}\cdot \frac{\partial f_{ex}}{\partial {\bf r}}-\nabla\Phi_{ex}\cdot \frac{\partial f_{ex}}{\partial {\bf v}}=\xi \frac{\partial}{\partial {\bf v}}(f_{ex}{\bf v}).
\end{eqnarray}
This exact equation, where $f_{ex}$ is expressed in terms of Dirac
functions, contains exactly the same information as the deterministic
equations (\ref{tz7})-(\ref{tz8}). For $\xi=0$, it reduces to the
Klimontovich equation of plasma physics.

\section{The case $N=2$ in one dimension}
\label{sec_ud}

In this Appendix, we study the statistical
mechanics of two particles in gravitational interaction in $d=1$ where
explicit results can be obtained.

\subsection{The microcanonical ensemble}
\label{sec_udm}

In $d=1$, the Hamiltonian of two particles in gravitational interaction can be written 
\begin{eqnarray}
\label{udm1}
H=\frac{1}{2}MV^{2}+\frac{1}{2}\mu v^{2}+G m_{1}m_{2} |r|,
\end{eqnarray}
where we have used the coordinates $(R,V)$ of the center of mass and
the coordinates $(r,v)$ of the reduced particle.  We note that the
energy takes only positive values. Introducing $x_{1}=(M/2)^{1/2}V$
and $x_{2}=(\mu/2)^{1/2}v$, the surface of phase space with energy
less than $E$ can be rewritten
\begin{eqnarray}
\label{udm2}
\Gamma(E)=\frac{8 R}{\sqrt{M \mu}}\int_{0}^{r_{m}} d{r}\int_{\| {\bf x}\|\le \sqrt{E-G m_{1}m_{2}r}}d^{2}{x}.
\end{eqnarray}
The last integral represents the surface of a disc with radius $\sqrt{E-G m_{1}m_{2} r}$. Thus, we get
\begin{eqnarray}
\label{udm3}
\Gamma(E)=\frac{8\pi R}{\sqrt{M \mu }}\int_{0}^{r_{m}} (E-G m_{1}m_{2}r) dr.
\end{eqnarray}
The range of integration $r_{m}$ is such that $E-G m_{1}m_{2}r\ge 0$
and $r\le R$. Therefore, $r_{m}=E/Gm_{1}m_{2}$ if $E\le G m_{1}m_{2}
R$ and $r_{m}=R$ if $E\ge G m_{1}m_{2} R$. The density of states in
the microcanonical ensemble is $g(E)=d\Gamma/dE$. Introducing
\begin{eqnarray}
\label{udm4}
A=\frac{8\pi}{\sqrt{M\mu}}, \quad \epsilon=\frac{E}{Gm_{1}m_{2}R}, \quad t=\frac{k_{B}T}{Gm_{1}m_{2}R},
\end{eqnarray}
we get
\begin{eqnarray}
\label{udm5}
g(\epsilon)=A R r_{m},
\end{eqnarray}
with $r_{m}=\epsilon R$ if $0\le \epsilon\le 1$ and $r_{m}=R$ if $\epsilon\ge 1$. 

\begin{figure}
\centering
\includegraphics[width=8cm]{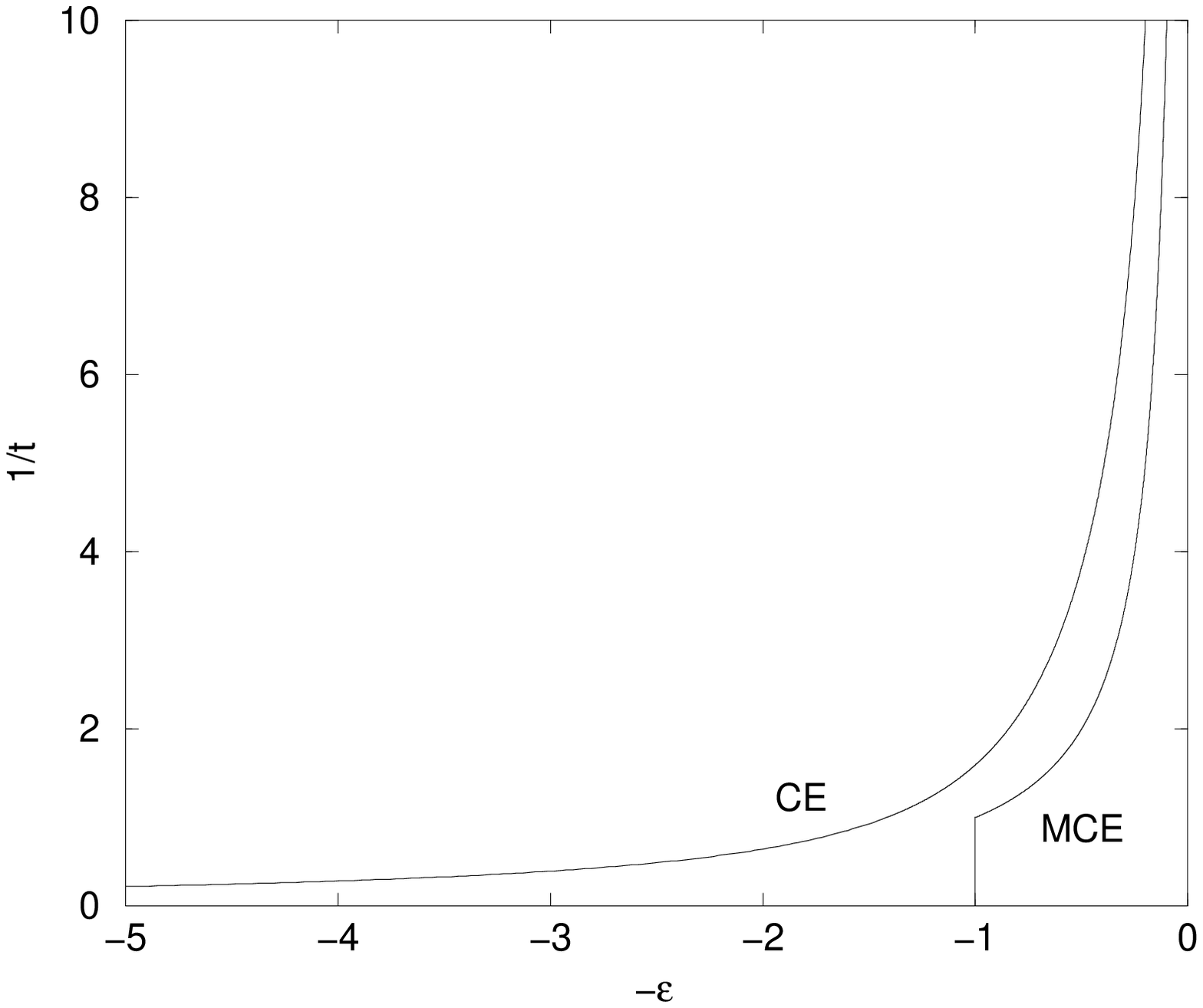}
\caption{Caloric curve in microcanonical (MCE) and canonical (CE) ensembles for the binary star model $N=2$ in one dimension. The caloric curves are similar in the two ensembles (without negative specific heat region) so that there is no phase transition contrary to the situation in $d=3$ \cite{paddy,rev}. There exists equilibrium states for all accessible energies $E\ge 0$ in MCE and for all temperatures in CE. It can be of interest to compare these curves  with the caloric curve obtained within the mean field approximation valid for $N\gg 1$ which presents a similar behaviour \cite{sc}.    }
\label{caloricd1}
\end{figure}

The caloric curve $t=t(\epsilon)$ and the equation of state in the
microcanonical ensemble can be obtained from
Eqs. (\ref{dm9})-(\ref{dm11}) and the exact expression (\ref{udm5}) of
the density of states. For $0\le \epsilon\le 1$, we obtain
\begin{eqnarray}
\label{udm6}
g(\epsilon)=AR^{2}\epsilon,
\end{eqnarray}
\begin{eqnarray}
\label{udm7}
k_{B}T=E\quad \rightarrow t=\epsilon,
\end{eqnarray}
\begin{eqnarray}
\label{udm8}
PV=2k_{B}T.
\end{eqnarray}
For $\epsilon\ge 1$, we obtain
\begin{eqnarray}
\label{udm9}
g(\epsilon)=AR^{2},
\end{eqnarray}
\begin{eqnarray}
\label{udm10}
k_{B}T=+\infty\quad \rightarrow t=+\infty,
\end{eqnarray}
\begin{eqnarray}
\label{udm11}
PV=2k_{B}T=+\infty.
\end{eqnarray}
The caloric curve in the microcanonical ensemble is represented in Fig. \ref{caloricd1}.

\subsection{The canonical ensemble}
\label{sec_udc}

For $N=2$ particles in gravitational interaction, the partition
function in the canonical ensemble can be written
\begin{eqnarray}
\label{udc1}
Z=\int e^{-\beta M\frac{V^{2}}{2}}e^{-\beta \mu\frac{v^{2}}{2}}
e^{-\beta m_{1}m_{2}G|r|}d{R}d{V}d{r}d{v}.\nonumber\\
\end{eqnarray}
Integrating over the velocities and over the position of the center of
mass\footnote{In fact, the domain of integration of ${r}$ depends on
${R}$. Therefore, our treatment (see also Sec. \ref{sec_n2}) is
approximate and will be improved in a future work.}, we get
\begin{eqnarray}
\label{udc2}
Z=\frac{8\pi R}{\beta \sqrt{M\mu}}\int_{0}^{R} 
e^{-\beta G m_{1}m_{2}r}\ dr.
\end{eqnarray}
Therefore, the partition function is given by
\begin{eqnarray}
\label{udc3}
Z=\frac{8\pi R}{\beta^{2} G (m_{1}m_{2})^{3/2}}\left (1-e^{-\beta Gm_{1}m_{2}R}\right ).
\end{eqnarray}

\begin{figure}
\centering
\includegraphics[width=8cm]{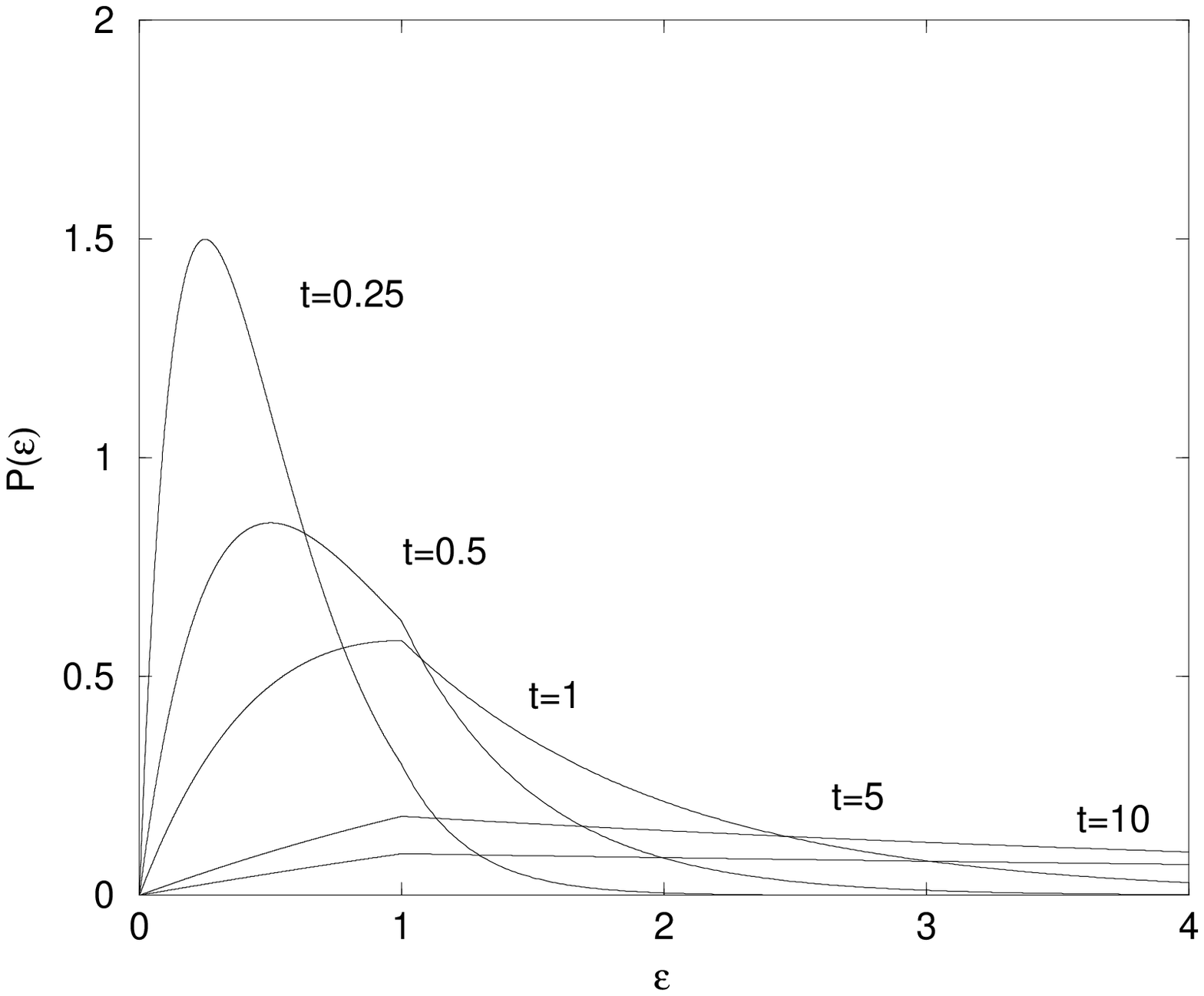}
\caption{ Distribution of energies for the binary star model $N=2$ in one dimension in the canonical ensemble. We have plotted this distribution for different values of the temperature.   }
\label{prob1}
\end{figure}

The pressure can be computed from Eq. (\ref{ca5}) leading to
\begin{eqnarray}
\label{udc4}
PV=k_{B}T\left\lbrack 1+\frac{Gm_{1}m_{2}\beta R}{e^{\beta G m_{1}m_{2}R}-1}\right\rbrack.
\end{eqnarray}
For $T\rightarrow +\infty$, we have $PV=2k_{B}T$ and for $T\rightarrow
0$, we have $PV=k_{B}T$ (this is the pressure created by an effective
{\it single} particle resulting from the collapse of the two
particles). The average energy obtained from Eqs. (\ref{ddc9}) and
(\ref{udc3}) is
\begin{eqnarray}
\label{udc5}
\langle\epsilon\rangle=2t-\frac{1}{e^{1/t}-1}.
\end{eqnarray}
It behaves like $\langle\epsilon\rangle\sim 2t$ for $t\rightarrow 0$
and like $\langle\epsilon\rangle\sim t$ for $t\rightarrow
+\infty$. The caloric curve in the canonical ensemble is represented
in Fig. \ref{caloricd1} and it is compared with the microcanonical
caloric curve. Finally, the energy distribution at temperature $T$ in
the canonical ensemble obtained from Eqs. (\ref{ddc11}), (\ref{udc3})
and (\ref{udm5}) is given by
\begin{eqnarray}
\label{udc6}
P(\epsilon)=\frac{\epsilon e^{-\epsilon/t}}{t^{2}(1-e^{-1/t})}, \qquad (0\le \epsilon\le 1),
\end{eqnarray}
\begin{eqnarray}
\label{udc7}
P(\epsilon)=\frac{e^{-\epsilon/t}}{t^{2}(1-e^{-1/t})}  \qquad (\epsilon\ge 1).
\end{eqnarray}
The distribution of energies is represented in Fig. \ref{prob1} for
different temperatures.

\end{document}